\documentclass{article}
\usepackage{appendix}
\usepackage{subfiles}
\usepackage[hidelinks]{hyperref}

\usepackage[
    maxcitenames=2,
    maxbibnames=3,
    style=authoryear,
    backend=biber
]{biblatex}

\newif\ifincludesupp   
\includesupptrue

\usepackage{xr-hyper} 
\usepackage[cmex10]{amsmath}
\usepackage{amsthm,amssymb}

\usepackage[pdftex]{graphicx} 
\graphicspath{{./figures/}}

\usepackage{bm}
\usepackage{subcaption}

\usepackage[a4paper, left=1in, right=1in, top=1in, bottom=1in]{geometry}

\usepackage{authblk}

\usepackage{xr} 
\usepackage{subfiles}

\newcommand{\fref}[1]{Figure~\ref{#1}}   
\newcommand{\cref}[1]{Chapter~\ref{#1}}  
\newcommand{\data}{\mathcal{D}}
\newcommand{\xvec}{\mathbf{x}}
\newcommand{\wvec}{\mathbf{w}}
\newcommand{\pmodel}{p_{\wvec}}
\newcommand{\wpost}{p(\wvec \mid \data)}
\newcommand{\opinion}{\pmodel(y \mid \xvec)}
\newcommand{\ens}{p_{\text{ens}}}

\newcommand{\au}{\text{AU}}
\newcommand{\eu}{\text{EU}}

\newcommand{\pmodern}{\textbf{\textit{premodern}}}
\newcommand{\modern}{\textbf{\textit{modern}}}
\newcommand{\scenario}{\textbf{\textit{scenario}}}

\date{\small Published in \textit{Machine Learning{:} Earth}}

\title{\Large \textbf{Am I Confused or Is This Confusing?:\\  Deep Ensembles for ENSO Uncertainty Quantification}\\[1em]}

\author[1]{Devin M. McAfee\thanks{Corresponding author: dmcafee@bu.edu}}
\author[2,3]{Elizabeth A. Barnes}

\affil[1]{Department of Atmospheric Science, Colorado State University}
\affil[2]{Faculty of Computing and Data Sciences, Boston University}
\affil[3]{Department of Earth and Environment, Boston University}

\begin{document}

\maketitle

\begin{abstract}
    Faithful uncertainty quantification (UQ) is paramount in high stakes climate prediction. Deep ensembles, or ensembles of probabilistic neural networks, are state of the art for UQ in machine learning (ML) and are growing increasingly popular for weather and climate prediction. However, detailed analyses of the mechanisms, strengths, and limitations of ensembles in these complex problem settings are lacking. We take a step towards filling this gap by deploying deep ensembles for predictability analysis of the El-Niño Southern Oscillation (ENSO) in the Community Earth System Model 2 Large Ensemble (CESM2-LE). Principally, we show that epistemic uncertainty, modeled by ensemble disagreement, robustly signals predictive error growth associated with shifts in the distributions of monthly sea-surface temperature (SST), ocean heat content (OHC), and zonal surface wind stress ($\tau_x$) anomalies under a climate change scenario. Conversely, we find that aleatoric uncertainty, which remains a popular measure of model confidence, becomes less reliable and behaves counterintuitively under climate-change-induced distributional shift. We highlight that, because ensemble performance improvement relative to the expected single model scales with epistemic uncertainty, ensemble improvement increases with distributional shift from climate change. This work demonstrates the utility of deep ensembles for modeling aleatoric and epistemic uncertainty in ML climate prediction, as well as the growing importance of robustly quantifying these two forms of uncertainty under anthropogenic warming.
\end{abstract}

\section{Introduction} 
\label{sec:intro}

Climate is \textit{confusing}. A chaotic web of nonlinear interactions between the ocean, atmosphere, and land, Earth's climate is, in many ways, random. The random and confusing nature of the climate system creates aleatoric uncertainty, i.e., unpredictability \parencite{hullermeier}, in the lagged relationships between climate variables. This leads to inevitable error in climate prediction, even if provided optimal model solutions.

In many heavily investigated climate prediction problems, aleatoric uncertainty is dependent on the input state \parencite{mariotti}. For instance, consider the El Niño-Southern Oscillation (ENSO). ENSO describes quasi-periodic variations in tropical Pacific sea-surfaces temperatures (SSTs), and the coupled atmospheric Walker circulation, on seasonal-to-multiannual timescales. Whether ENSO is in a warm (El Niño) or cool (La Niña) phase has significant influence beyond just the Pacific basin, and modulates the global atmospheric circulation, as well as global surface temperatures \parencite{mcphadenenso}. The ENSO signal thereby influences aleatoric uncertainty across a range of climate prediction tasks \parencite{mariotti}.

The existence of state-dependent aleatoric uncertainty, e.g., as forced by ENSO, implies that the risk of deterministic climate prediction depends on input. A trustworthy climate forecasting model must robustly quantify the uncertainty of its predictions so that users can anticipate this risk \parencite{gneiting}. Indeed, uncertainty quantification (UQ) has long been integral to weather and climate prediction \parencite{murphyhistory}, leading to, e.g., the development of ensemble forecasting \parencite{epsteinensemble, leith, lewisRoots}, data assimilation \parencite{kalnay}, and widely used forecast verification methods \parencite{brier, epsteinscoring, murphyreliability, brocker, palmerss}. 

In the past decade, deep neural networks have gained significant traction in the world of UQ. Modern neural networks designed for UQ can achieve impressive predictive capacity and reliability \parencite{minderer}, and have shown promise across scientific disciplines \parencite{abdarReview}. In the climate domain, probabilistic neural networks, often of modest depth, are an emerging tool for skillful state-dependent aleatoric uncertainty estimation \parencite{mayer, luoclimatebayes, delaunay, gordonunc}. However, in such applications, aleatoric uncertainty estimates are commonly conflated with model confidence, despite the fact that neural networks can predict low outcome uncertainty while unconfident and $\textit{confused}$  \parencite{galdropout}. Conversely, a skilled model will confidently predict high outcome uncertainty when input conditions are inherently uninformative. For instance, seasonal ENSO prediction is canonically more difficult through boreal spring \parencite{barnston}. This deficiency is shared among all ENSO prediction models, both physical and statistical \parencite{ehsan}. Thus, if an ENSO prediction model is less accurate during spring, this does not imply that the model is confused and underconfident, or that there exists a different model with a more informed answer.  Therefore, in climate prediction, it is useful to quantify a model's confidence separate from its aleatoric uncertainty estimates \parencite{kendall, hullermeier}. 

Accurately gauging model confidence is particularly important for prediction under covariate shift. Covariate shift occurs when the input distribution changes at testing relative to training, invariably degrading model performance \parencite{ovadia, hullermeier}.  The most societally relevant mechanism of covariate shift affecting the present Earth system is anthropogenic climate change. By perturbing mechanisms of internal climate variability, climate change has cascading effects across time scales \parencite{ipccch3}. In this work, we demonstrate how shifts in the monthly variability of sea-surface temperature (SST), upper ocean heat content (OHC), and zonal wind stress ($\tau_x$) induced by climate change impact ENSO prediction. Specifically, we train deep ensembles -- ensembles of probabilistic neural networks -- to predict ENSO in large climate model simulations, and identify decreased predictive skill under a future warming scenario due to projected shifts in input variability. We describe how deep ensembles mitigate risk during these shifts, relative to single models, by accounting for the epistemic uncertainty about their predictions, which we interpret as model confidence. We find that epistemic uncertainty, quantified as ensemble disagreement, provides a coherent signal of climate-change-induced shift, even when the aleatoric uncertainty signal is muted. Along these lines, we argue for deep ensembles as a simple yet robust tool for climate UQ, enabling users to gauge both aleatoric and epistemic uncertainty in such high-stakes prediction tasks.

We organize the manuscript as follows. In section \ref{sec:deepensembles}, we summarize the literature on deep ensembles, which are a state of the art for UQ in ML. In section \ref{sec:decomp}, we describe metrics for quantifying aleatoric and epistemic uncertainty. In section \ref{sec:experiments}, we describe the ENSO prediction task examined in this study. In section \ref{sec:ensemble_improvement}, we show that, by accounting for epistemic uncertainty, deep ensembles outperform single models. In section \ref{sec:id}, we illustrate the utility of deep ensemble aleatoric uncertainty estimates in characterizing established features of ENSO predictability, for inputs within the training distribution. In section \ref{sec:shift}, we leave the training distribution, and analyze how projected changes in tropical Pacific variability impact estimations of aleatoric and epistemic uncertainty in deep ensembles. We conclude with a discussion regarding the importance of modeling epistemic uncertainty as climate changes. 

\section{Deep Ensembles}
\label{sec:deepensembles}

The standard approach for quantifying state-dependent aleatoric uncertainty in ML is to treat the target $Y$ as a random variable generated by an underlying probability distribution $p^{*}(y \mid \xvec)$ conditioned on predictors $\xvec$. A probabilistic forecasting model $\pmodel(y \mid \xvec)$ is an approximation of $p^{*}(y \mid \xvec)$, with weights $\wvec$ learned from a training dataset $\data = \{(\xvec_1,y_1),\dots,(\xvec_N, y_N)\}$ sampled from $p^{*}$. In complex climate prediction, all models are wrong.  While neural networks are useful for modeling these high-dimensional problems, it is highly unlikely a selected architecture can perfectly represent $p^{*}$ or, even if so, that gradient descent would find the optimal solution \parencite{blasiokproper, wildlink}. Therefore, the predictions of trained neural networks are influenced by model uncertainty. 

Typically, optimization is performed for a fixed neural network architecture, so, rather than representing model uncertainty completely, we instead consider uncertainty in the model weights $\wvec$. The Bayesian practice is to describe weight uncertainty through the posterior distribution $\wpost$. Each weight vector in the support of the posterior represents a subjective opinion about the data's aleatoric uncertainty $\opinion$. The posterior describes the relative importance of opinion, determined by ability to generate the training data and prior belief. 

\begin{equation}
    \wpost = \frac{p_{\wvec}(\data)p(\wvec)}{p(\data)}
    \label{bayesrule}
\end{equation}

The model evidence $p(\data)$ involves an integration over the weight space, making \eqref{bayesrule} computationally intractable for large neural networks. Therefore, for Bayesian deep learning, the posterior must be approximated using less expensive methods. A widely used technique is variational inference, which models the posterior as a simplified distribution; most commonly a Gaussian is assumed \parencite{blundell, blei}. The degree of simplification required to ensure tractability of variational inference is often too significant for this strategy to outperform standard deep ensembles in practice \parencite{ovadia, ashukha, izmailovhmc, wildlink}. This is because, in deep learning, posteriors are highly complex and multimodal \parencite{wilsonpersp}, an environment where deep ensembles thrive. 

Deep ensembles are groups of probabilistic neural networks trained independently on the same dataset. Generally, deep ensembles have a fixed architecture over their components\footnote{We use the terminology ``ensemble component'' for deep ensembles and reserve the more traditional ``ensemble member'' for the climate model ensemble used for training (see sec \ref{sec:cesm2})}, which only differ by the random seeds of their weight and training batch order initializations \parencite{leeMheads, lakshminarayanan}. At inference, deep ensembles predict a uniform mixture of component distributions.

\begin{equation}
    \ens(y \mid \xvec) = \frac{1}{M} \sum_{i=1}^{M} p_{\wvec_i}(y \mid \xvec)
    \label{empd}
\end{equation}

where $M$ is the ensemble size, $\ens(y \mid \xvec)$ is the deep ensemble's predictive distribution, and $p_{\wvec_i}(y \mid \xvec)$ is the distribution output by the i$^{\text{th}}$ ensemble component. 

\vspace{5pt}

Components are trained the standard way, by minimizing a proper loss function. Proper losses are standard because they are only minimized by the data-generating distribution itself, thereby incentivizing learners to faithfully report uncertainty \parencite{raftery}. The most popular example is the negative log-likelihood (NLL), as its minimization equates to maximum likelihood estimation (MLE).

\begin{equation}
    \text{NLL}(p, y) = -\log{p(y \mid \xvec)}
    \label{nll}
\end{equation}

Overfitting is a concern for MLEs of large neural networks, especially when the training data contains significant aleatoric uncertainty. Thus, we regularize with a prior $p(\wvec)$. The standard approach, and the one in this study, is to simply add an L2 penalty to the NLL loss, which assumes an isotropic Gaussian prior over weights, whose variance determines the regularization constant (see section \ref{sec:prior}). The isotropic Gaussian is also the standard prior in Bayesian deep learning \parencite{izmailovhmc}. The regularized loss is given by 

\begin{equation}
    L(\pmodel) = -\sum_{j=1}^{N} \log{p_{\wvec}(y_j \mid \xvec_j)} + \frac{1}{2\sigma^2}\|\mathbf{\wvec}\|_2^2
    \label{loss}
\end{equation}

where $N$ is the number of training samples and $\sigma^2$ is the prior variance. 

\begin{figure}
    \centering
     \includegraphics[width=0.7\textwidth]{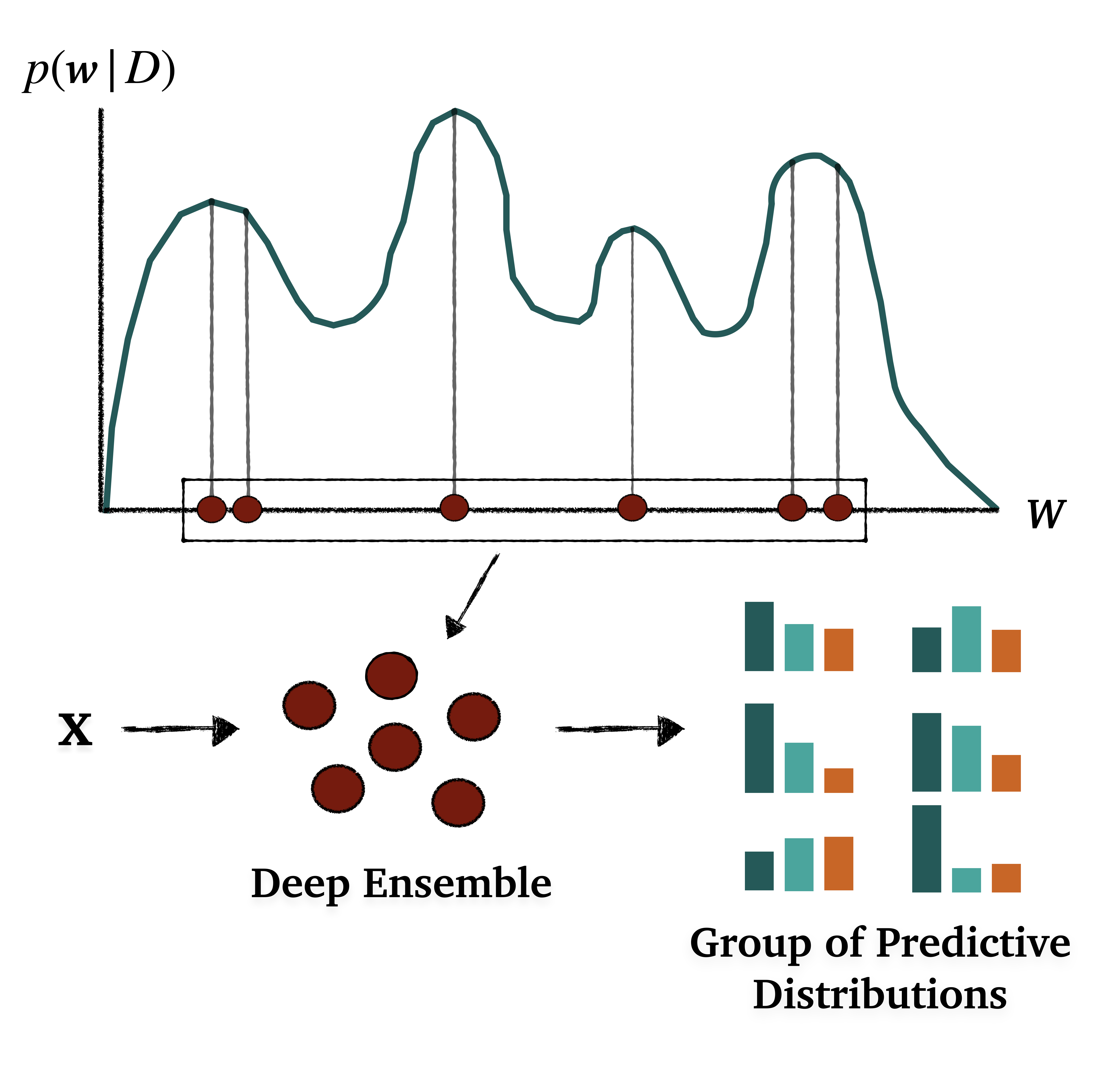}
    \caption{Schematic of the deep ensemble framework given a multimodal posterior for a three-class classification problem (probability masses denoted by colored bars). Each component is a potentially unique mode of $\wpost$, which produces potentially unique functional representations of the data.} 
    \label{Fig1}
\end{figure}

With the inclusion of a prior, the training loss becomes the negative log posterior density $-\log{\wpost}$ (see section \ref{sec:prior}). Hence, deep ensembles are collections of local modes of $\wpost$ \parencite{wilsonpersp, wildlink}, which we illustrate schematically in Fig. \ref{Fig1}, an adaptation of Fig. 3 from \textcite{wilsonpersp}. This has spurred debate in the literature over whether deep ensembles should be perceived as ``Bayesian" \parencite{wilsonpersp, dangelo, wildlink}. Since deep ensembles are mixtures of delta functions at posterior modes, \eqref{empd} is a discrete approximation of a Bayesian model average \parencite{wilsonpersp}. However, as components are restricted to lie within local modes, deep ensembles don't converge to the true posterior in the limit of infinite size \parencite{wildlink, dangelo}, so they are not Bayesian in the classical sense. Yet, by exploiting the multimodal structure of the posterior, deep ensembles outperform parametric approaches like variational inference \parencite{wilsonpersp, wildlink}. In fact, even principled ensembling strategies with theoretical convergence guarantees don't beat naive ensembles \parencite{dangelo, wildlink}, which may be a consequence of the bountiful number of explorable local modes in deep learning \parencite{wildlink, luoclimatebayes, fort}.

Perhaps the most important property of deep ensembles is that they are guaranteed to mitigate risk compared to training a single model. This is because proper losses are convex, so, by Jensen's inequality, the loss of the deep ensemble prediction is at most the expected component loss  \parencite{abePredictiveDiversity}. That is, for convex loss $\phi$, $\phi(\text{E}[p], y) \leq \text{E}[\phi(p, y)]$. Further, in practice, deep ensembles tend to generalize better than all of their individual components \parencite{nixonbootstrap, wangdisteq}, which is also observed in this work.

\section{Aleatoric and Epistemic UQ}
\label{sec:decomp}
For each input, we quantify an ensemble's aleatoric uncertainty estimate, $\au$, as the component-mean entropy. This approximates the conditional entropy $\text{H}(Y \mid \mathbf{W})$ \parencite{depeweg}, or the predictive uncertainty which remains given knowledge of the weights. We deal with classification problems in this work, where each component predicts the parameters of a categorical distribution over $\mathcal{Y} = \{1,\dots,K\}$. AU is expressed as 

\begin{equation}
    \au(\xvec) = -\frac{1}{M}\sum_{i=1}^{M}  \sum_{k=1}^{K} p_{\wvec_i}(y = k \mid \xvec) \log{p_{\wvec_i}(y = k \mid \xvec)}
    \label{au_eq}
\end{equation}

AU takes values in $[0, \log K]$. 

The weight uncertainty modeled by deep ensembles induces functional disagreement among components at each input \parencite{fort}. This disagreement captures epistemic uncertainty, corresponding to a range of plausible predictive distributions for a given input. We quantify epistemic uncertainty, $\eu$, using the following metric: 

\begin{equation}
    \eu(\xvec) =  \frac{1}{M} \sum_{k = 1}^{K} \sum_{i=1}^{M} (p_{\wvec_i}(y = k \mid \xvec) - \ens(y = k \mid \xvec))^2
    \label{eu_eq}
\end{equation}

$\eu$, which takes values in $[0, \frac{K-1}{K}]$, is the ensemble variance of class probabilities summed across classes. This variance-based definition of disagreement is popular in the literature and can be derived by decomposing the variance of $Y$ using the law of total variance \parencite{depeweg, duan, schreck}. Importantly, this metric equals the ensemble's inputwise improvement over it's mean-performing component, as measured by Brier score (BS) (\cite{abePredictiveDiversity}, see \ref{sec:brierimprov}): 

\[
[y=k] =
\begin{cases}
1 & \text{if } y=k, \\
0 & \text{otherwise.}
\end{cases}
\]

\begin{equation}
\text{BS}(p, y) = \sum_{k=1}^{K} 
\left( p(y=k \mid \xvec) - [y=k] \right)^2
\end{equation}

Brier score \parencite{brier} is a proper loss \parencite{raftery} and takes values in $[0,2]$. 

We emphasize that, unlike loss functions, $\au$ and $\eu$ are defined independently of the true label and thus depend solely on ensemble predictions. That is, being uncertainty measures, $\au$ and $\eu$ can be evaluated for unlabeled data, such as in operational forecasting.  Further, while defining aleatoric and epistemic uncertainty on different scales is nonstandard \parencite{depeweg, wimmer}, the magnitudes of $\au$ and $\eu$ are not compared in this study.


\section{Experiments}
\label{sec:experiments}
\subsection{Climate Model Data}
\label{sec:cesm2}

We use data from the Community Earth Systems Model 2 Large Ensemble (CESM2-LE) \parencite{cesm2, rodgers}. This dataset contains simulations from 100 members covering the period [1850, 2100], with historical forcings until 2014 and SSP3-7.0 forcings thereafter \parencite{rodgers}. We use monthly fields of sea-surface temperature (SST) and zonal surface wind stress ($\tau_x$) over all marine grid points equatorward of 45 degrees latitude. We define upper ocean heat content (OHC) as the vertically integrated ocean temperature to a depth of 300 meters, as is done in \textcite{ham}. 

We separate the training, validation, and testing datasets based on CESM2 member, which removes temporal dependencies within members. We ensure the training and testing datasets share no micro-perturbed initializations to mitigate the influence of ocean initial condition memory within the first few decades after initialization \parencite{rodgers, desersmile}. We split among CESM2 members [1, 50], which share biomass burning forcings \parencite{rodgers}. To ensure robustness of results, we use CESM2 members [1, 20] for training, [21, 25] for validation, and reserve [26, 50] for testing.

For each variable, we compute standardized monthly anomalies relative to the period [1850, 1949], and detrend by subtracting the CESM2-LE-mean anomaly (with respect to all 100 members) at each time step to produce inputs. Since all CESM2 members are constrained by the same external forcings, the mean of the CESM2-LE distribution estimates the forced response at a month in time \parencite{frankcombe}. 

We define an ENSO index using the first principal component of the internal variability of tropical Pacific SSTs over the period [1850, 1949] across all CESM2 members. We create categorical labels by quartile binning the ENSO indices. In ascending quartile order, the four classes are hereafter referred to as La Niña (LN), Cold Neutral (CN), Warm Neutral (WN), and El Niño (EN), respectively. Throughout the manuscript, we use the term ENSO ``phase'' to mean class index. 

\subsection{Problem Setup and Deep Ensemble Design}

\begin{figure}[h!]
    \centering
     \includegraphics[width=1\textwidth,trim=4 4 4 4, clip]{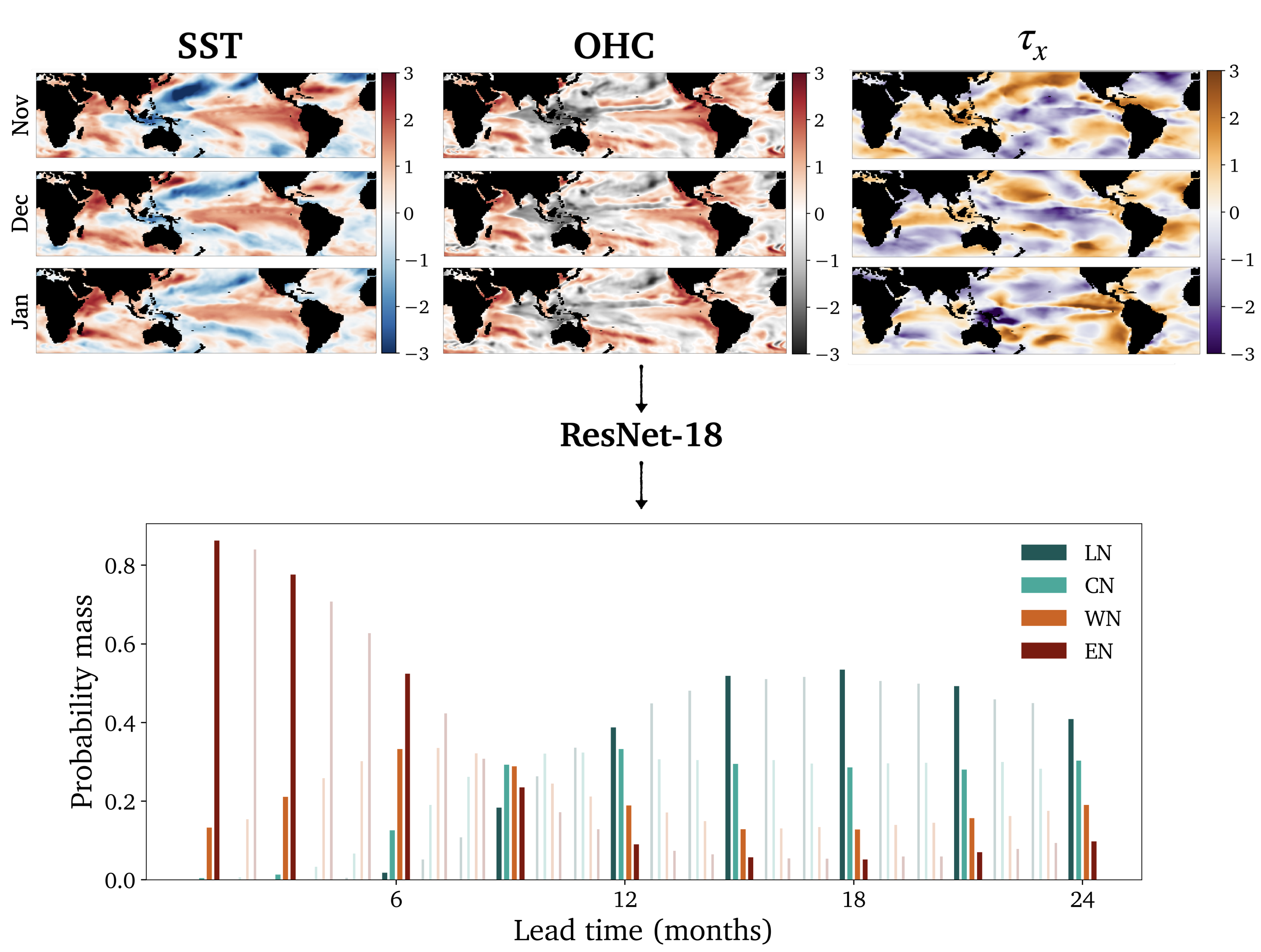}
     \caption{Illustration of the study's problem setup using an example prediction from a component of $\pmodern$ initialized in January. A ResNet-18 model ingests the last three months of anomalies (in units of standard deviations) and outputs categorical distributions over ENSO classes for the next 24 months. The highlighted distributions are for leads 1, 3, 6, $\dots$ and 24 months, and the faint distributions are for the remaining leads.}

    \label{setup}
\end{figure}

As illustrated in \fref{setup}, to examine the behavior of uncertainty across forecast lead time, we task each deep ensemble component to the predict the ENSO index over months ($t+1,\dots, t+24$ months) from SST, OHC, and $\tau_x$ during months ($t-2, t-1, t$). All metrics, including uncertainty measures and losses, are calculated separately for each lead time. Forecasting across multiple lead times aligns with operational practice and is thus standard for statistical ENSO prediction in the literature \parencite{ham, ehsan, chenENSO}. We train components to predict each lead time simultaneously, as done in, e.g., \textcite{wangsalinity}, instead of building a unique ensemble for every lead time, to facilitate large ensemble training. Our ensembles are considerably larger ($M=100$) than those typically trained for ENSO prediction to allow for robust comparison of component and ensemble generalization. We reproduce our main findings using the alternative fixed lead training approach with $M=10$ ensembles in section \ref{sec:fltd}, and discuss discrepancies between the two approaches where relevant.


We employ a modified ResNet-18 \parencite{heresnet} as the component architecture. ResNet-18 is a deep convolutional neural network, which makes use of skip connections to obtain a smoother loss surface, resulting in more stable optimization \parencite{lilandscape}. To the base ResNet-18 architecture, we prepend a nontunable circular padding layer, which pads a fixed number of zeros around each input channel to prevent dilution of features at the spatial domain edges, and a tunable reduction layer, a convolutional layer which compresses the nine input channels into three channels which can be input to ResNet-18. The output of ResNet-18 is fed into a fully-connected layer, which outputs 4-class logit vectors for each lead time. A softmax activation is then applied to obtain categorical parameters for each lead. 

We train three $M=100$ ensembles: 

\begin{enumerate}
    
    \item{The $\pmodern$ ensemble is trained with inputs from [1850, 1949].}

    \item{The $\modern$ ensemble is trained with inputs from [1850, 2024].}
    
    \item{The $\scenario$ ensemble is trained with inputs from [1850, 2098].}

\end{enumerate}

We perform inference with each ensemble for all inputs ([1850, 2098]) of the testing set. 

\section{Results}
\label{sec:results}
\subsection{Ensemble Improvement}
\label{sec:ensemble_improvement}

\begin{figure}[h!]
    \hspace{-0.3cm}
    \centering
     \includegraphics[width=0.95\textwidth,trim=4 4 4 4, clip]{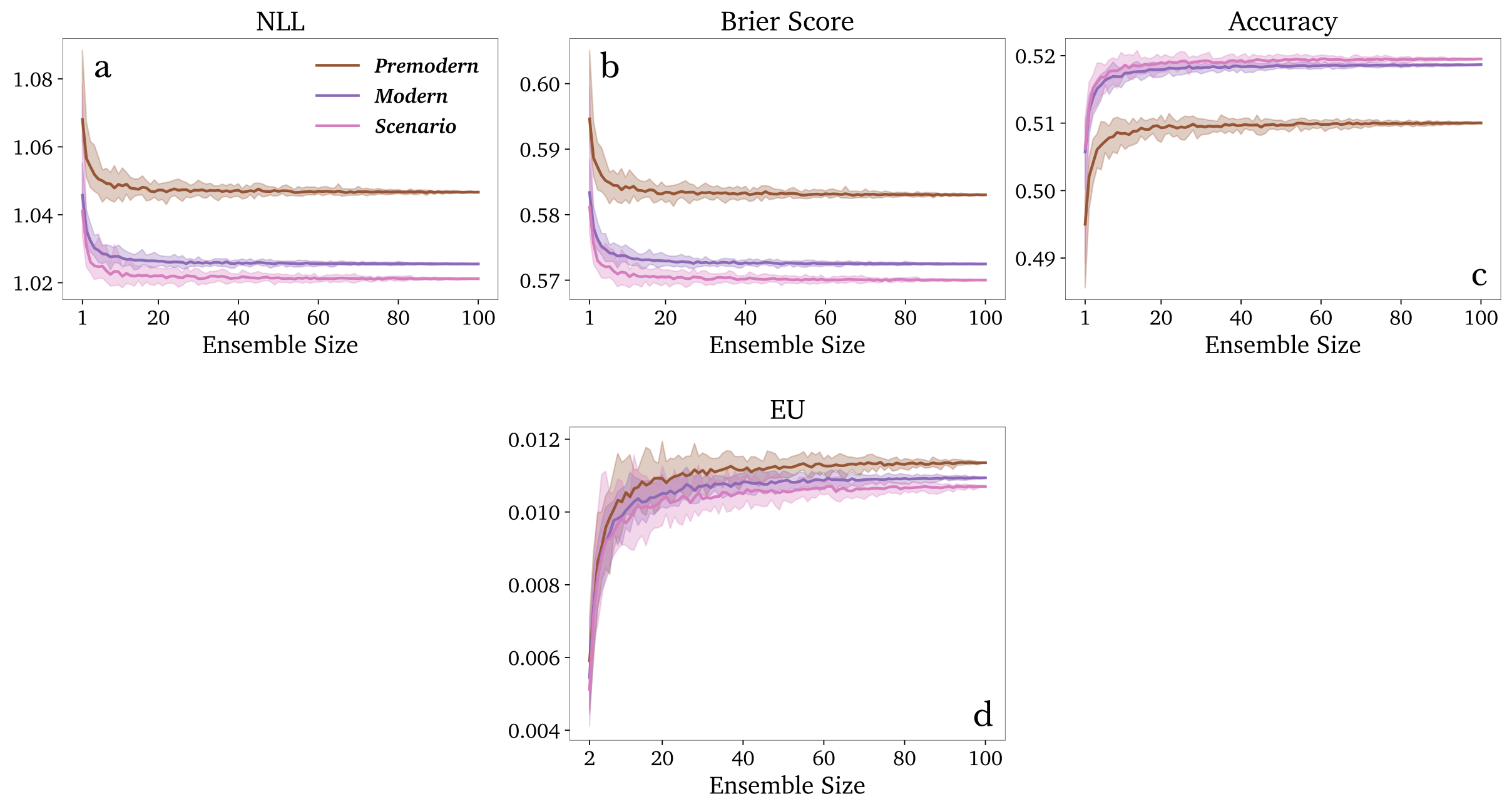}
     \caption{(a-c) Performance and (d) EU for the period [1850, 1949], averaged across leads, as a function of ensemble size for $\pmodern$, $\modern$, and $\scenario$. Shadings cover the range of scores from random subsampling of ensemble components without replacement, and curves represent the mean scores across subsamples.}
    \label{capacity}
\end{figure}

In this section, we analyze the performance of $\pmodern$, $\modern$, and $\scenario$ on the premodern period [1850, 1949] of the testing set, for in-distribution performance comparison across the ensembles. The remainder of the testing time series is reserved for the covariate shift analysis in section \ref{sec:shift}. Unless otherwise stated, all temporal averages are also averaged across CESM2 testing members.

The skill of each ensemble improves with increasing size, as shown in \fref{capacity}, which is a canonical property of deep ensembles \parencite{lakshminarayanan, wangdisteq}. Importantly, this performance improvement aligns with increases in $\eu$, indicating that, unsurprisingly, the ensembles benefit from incorporating epistemic uncertainty in their predictions. UQ performance and accuracy (of the top-label prediction) improve up to $\sim$ 10-20 components, where skill plateaus. $\scenario$ outperforms $\modern$, which outperforms $\pmodern$, because increasing the volume of training data decreases epistemic uncertainty. This is shown in \fref{capacity}d, where EU decreases from $\pmodern$ to $\modern$ to $\scenario$ across ensemble sizes. Since EU measures ensemble improvement, this suggests that the value of ensembling increases as data becomes limited.


\begin{figure}[h!]
    \centering
     \includegraphics[width=0.9\textwidth]{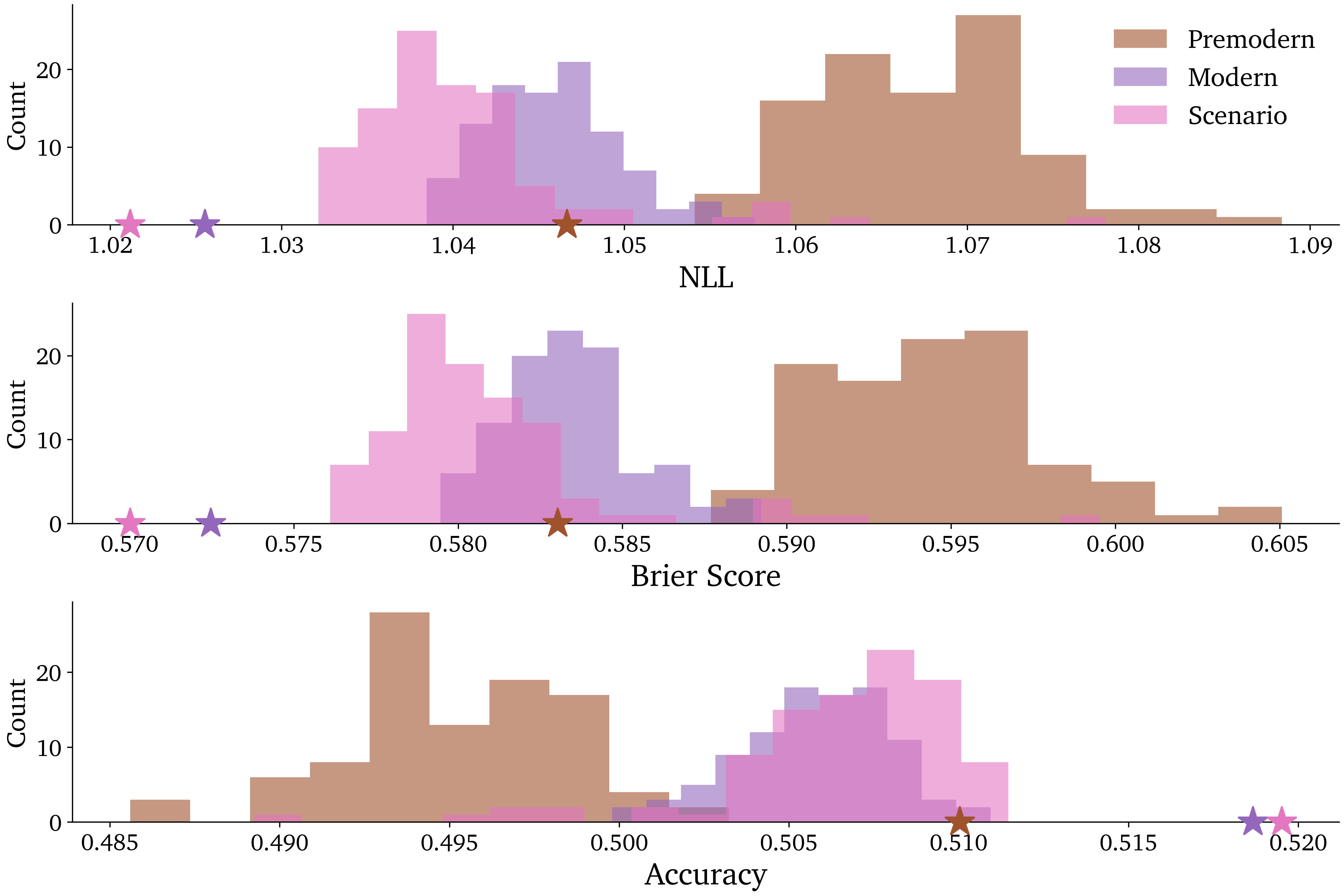}
     \caption{Mean testing scores, averaged across leads, for ensemble components (histograms) and deep ensembles (stars) over the premodern period.}
    \label{improvement}
\end{figure}

\begin{figure}[h!]
    \centering
     \includegraphics[width=0.8\textwidth]{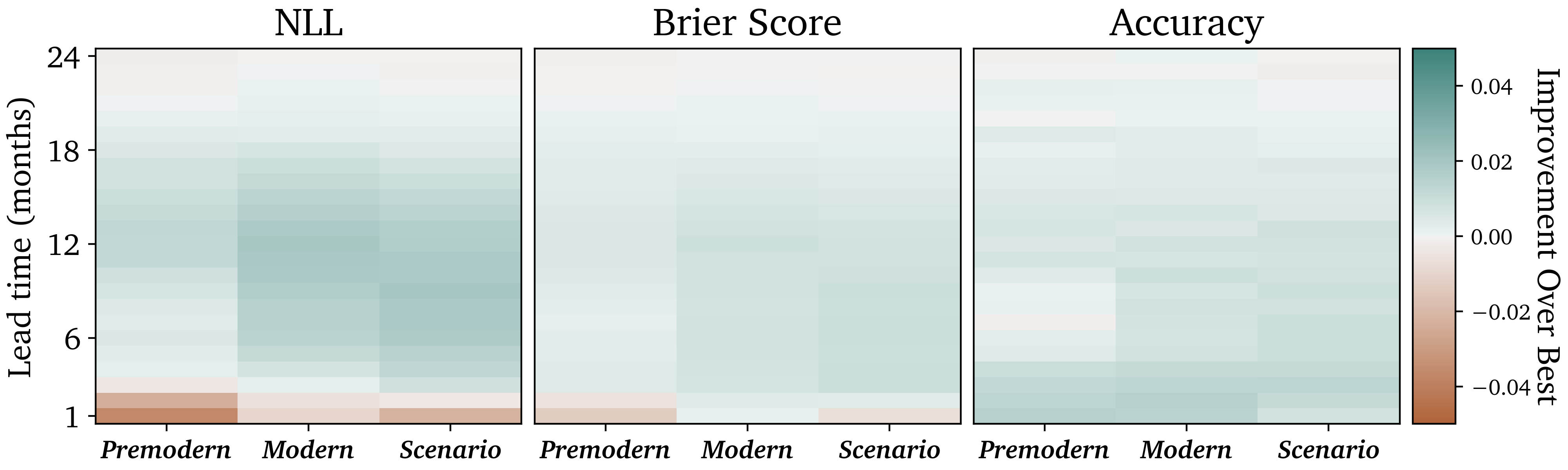}
     \caption{Difference in performance between ensemble and best performing component on the testing set over the premodern period.}
    \label{ens_vs_best}
\end{figure}

\fref{improvement} shows that each ensemble outperforms all of its components when averaged over lead times. As shown in \fref{ens_vs_best}, this ensemble improvement is present across lead times in accuracy but, surprisingly, not in NLL or Brier Score, particularly for leads within 3 months. We do not observe this contradiction when using the fixed lead approach, indicating this result is partially an artifact of training interference from longer, noisier leads (see \fref{ens_vs_best_flt}). However, the discrepancy between the improvement in accuracy and worsening of NLL suggests that ensemble predictions are miscalibrated relative to their components at short leads, and this miscalibration is also observed in the fixed lead scheme. We discuss this calibration bias in greater depth in section \ref{sec:clb}. In brief, it is a consequence of (1) the ensembles' improved accuracy over all components, (2) ensemble predictive entropy being upper bounded by AU, and (3) components being well calibrated within distribution \parencite{rahaman, wu}. In essence, an ensemble's predictions cannot sharpen to compensate for its improved accuracy, yielding miscalibration unless the components themselves are miscalibrated, which is common when optimizing for accuracy instead of proper loss \parencite{guo}.  We highlight that deep ensembles are not uniquely vulnerable to this artifact and that it generalizes to all UQ methods which improve accuracy by averaging over near-calibrated predictions, such as variational inference or Bayesian ensemble methods. Fortunately, this calibration bias, and its negative impacts on NLL and Brier Score at short leads, is effectively corrected by applying temperature scaling to the ensemble predictive distributions \parencite{rahaman} (see Figures \ref{reliability_scaled} and \ref{ens_vs_best_scaled}).

\subsection{AU In-Distribution}

\label{sec:id}

\begin{figure}
    \centering
     \includegraphics[width=0.9\textwidth]{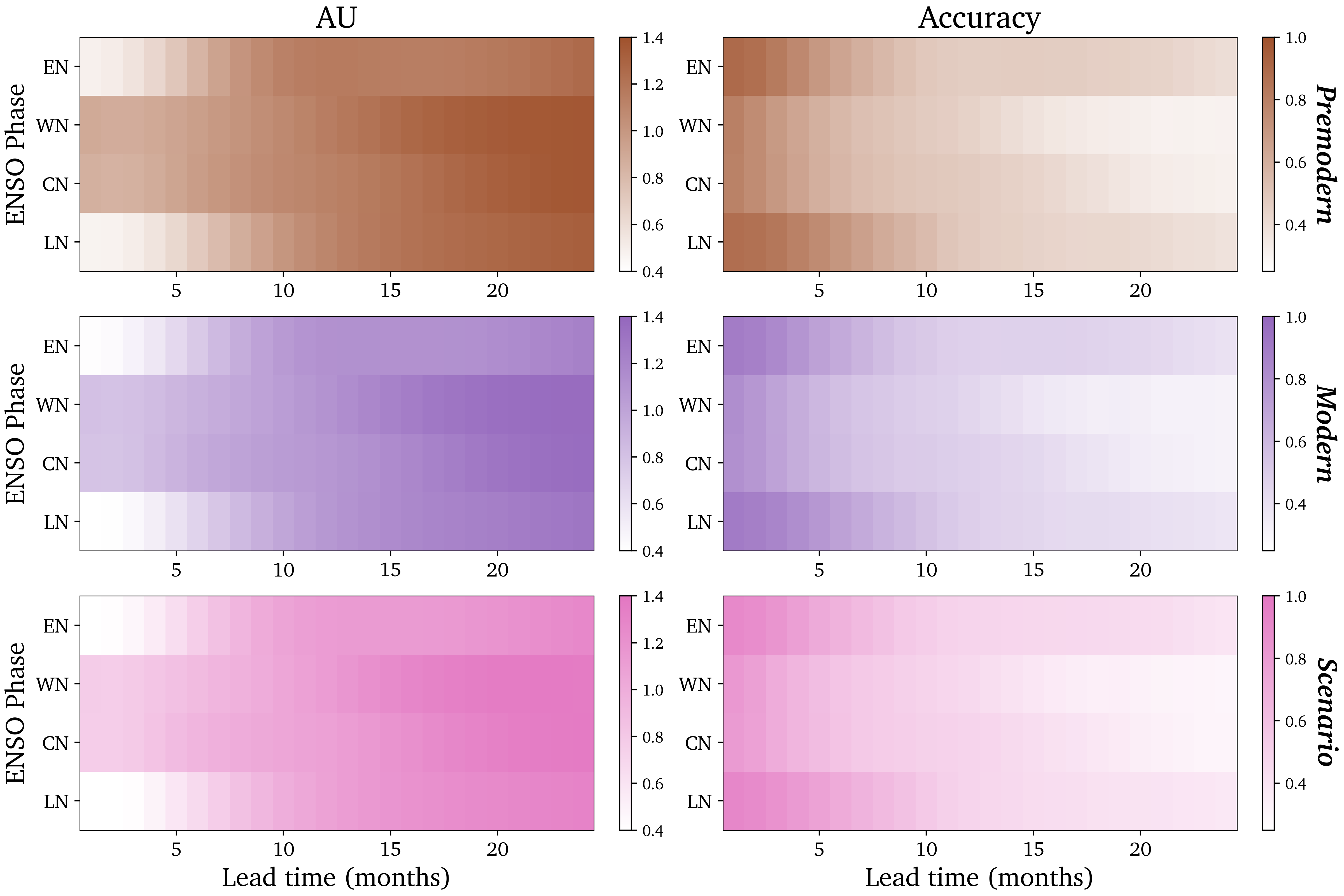}
     \caption{Mean $\au$ and accuracy as a function of input ENSO phase and lead time.}
    \label{unc_phase}
\end{figure}

For inputs near the training distribution, a trustworthy climate UQ model should produce state-dependent aleatoric uncertainty estimates which align with established patterns of climate predictability. In this section, we qualitatively assess the ability of AU from each ensemble in diagnosing known features of ENSO predictability for testing inputs in the premodern period. In the following section, we contrast this with the degradation of AU as inputs shift away from the training distribution due to climate change.  

\fref{unc_phase} shows the variability of AU across input ENSO conditions, represented by the ENSO phase at prediction time. Each ensemble is roughly equivalent in its in-distribution AU characteristics. AU identifies a rapid decrease in predictability with lead time, which is a universal property of chaotic climate prediction. Because the components are well calibrated (see section \ref{sec:clb}), accuracy rises as AU decreases with shortening lead time. The ensembles are considerably more accurate than chance (which, for quartile prediction, is also climatology) for all lead times, maintaining at least $34.9\%$ accuracy for two-year prediction in the premodern period. ENSO evolution is least predictable from neutral conditions which are characterized by muted Pacific SST anomalies and weakened ocean-atmosphere coupling compared to the strong phases.  For leads within one year, AU is lower and decreases more slowly with increasing lead for LN inputs than EN inputs.  This is consistent with the higher instability, i.e., shorter persistence timescale, of El Niño relative to La Niña \parencite{okumura}. Conversely, AU plateaus between leads $\sim$ 12 and 18 months for EN inputs. As a result, beyond one year, predictions from EN inputs have lower AU than those from LN inputs. The lag autocorrelation of ENSO-related Pacific anomalies falls to zero at 12 months in CESM2 \parencite{capotondi}, after which uncertainty related to phase transition is maximal. Hence, the observed difference in long-range AU between EN and LN inputs aligns with existing evidence that the transition from El Niño to La Niña is more predictable than from La Niña to El Niño. This asymmetry is tied to stronger meridional ocean heat transport--and induced OHC anomalies--in the western tropical Pacific following mature El Niño than La Niña, which is a key precursor of ENSO transition \parencite {planton, sharmila} (see \fref{att}).

\begin{figure}
    \centering
     \includegraphics[width=0.95\textwidth]{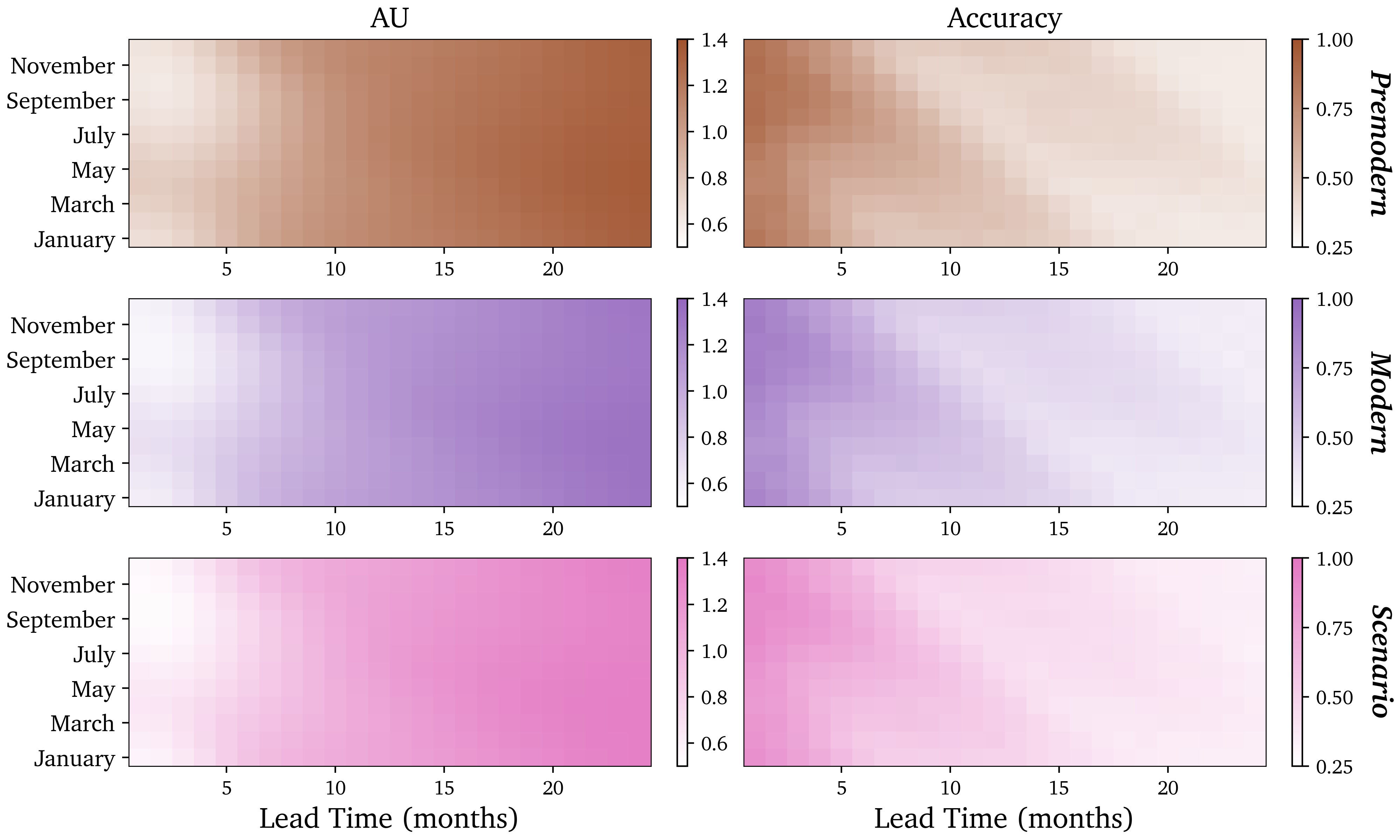}
     \caption{Mean AU and accuracy as a function of forecast month and lead time.}
    \label{seasonality}
\end{figure}

In \fref{seasonality}, we show the seasonal variability of AU and accuracy. Within $\sim$ 5 months, ensemble accuracy is maximal for forecasts beginning in boreal winter, during which AU is appropriately minimal. This is a known characteristic of ENSO predictability and is primarily due to seasonal phase locking, or the tendency for ENSO events to peak during boreal winter \parencite{chenphaselockingcesm2}, enhancing short-range predictability. Decreased AU for winter-initialized forecasts extends out a few months in lead time before meeting the spring predictability barrier. In \fref{seasonality}, the sharp diagonal gradient in AU and accuracy, within roughly 10 months, associated with forecasts through boreal spring is the classic signal of the spring predictability barrier \parencite{barnston}. This barrier is partially a consequence of the higher prevalence of neutral conditions outside of winter due to phase locking but still remains after conditioning on neutral phase inputs (see \fref{seasonality_phase}), potentially due to stochastic forcing from westerly wind bursts \parencite{lopezkirtman}, in addition to model deficiencies. Accuracy increases with lead time in parts of \fref{seasonality}, a peculiar feature not unique to our ENSO model \parencite{lounewmann, chenENSO}. Specifically, for forecasts initialized during El Niño in boreal winter, ensemble AU (accuracy) rapidly increases (decreases) through the following spring, due to the spring predictability barrier. AU then plateaus before decreasing into the following autumn and winter, coinciding with increasing accuracy, due to phase locking and the aforementioned predictability of El Niño-to-La Niña transition (see \fref{nov_en_au_acc}).

\subsection{AU and EU under Climate Change Shift}
\label{sec:shift}

\begin{figure}[h!]
    \hspace*{-1cm} 
    \centering
     \includegraphics[width=0.85\textwidth,trim=4 4 4 4, clip]{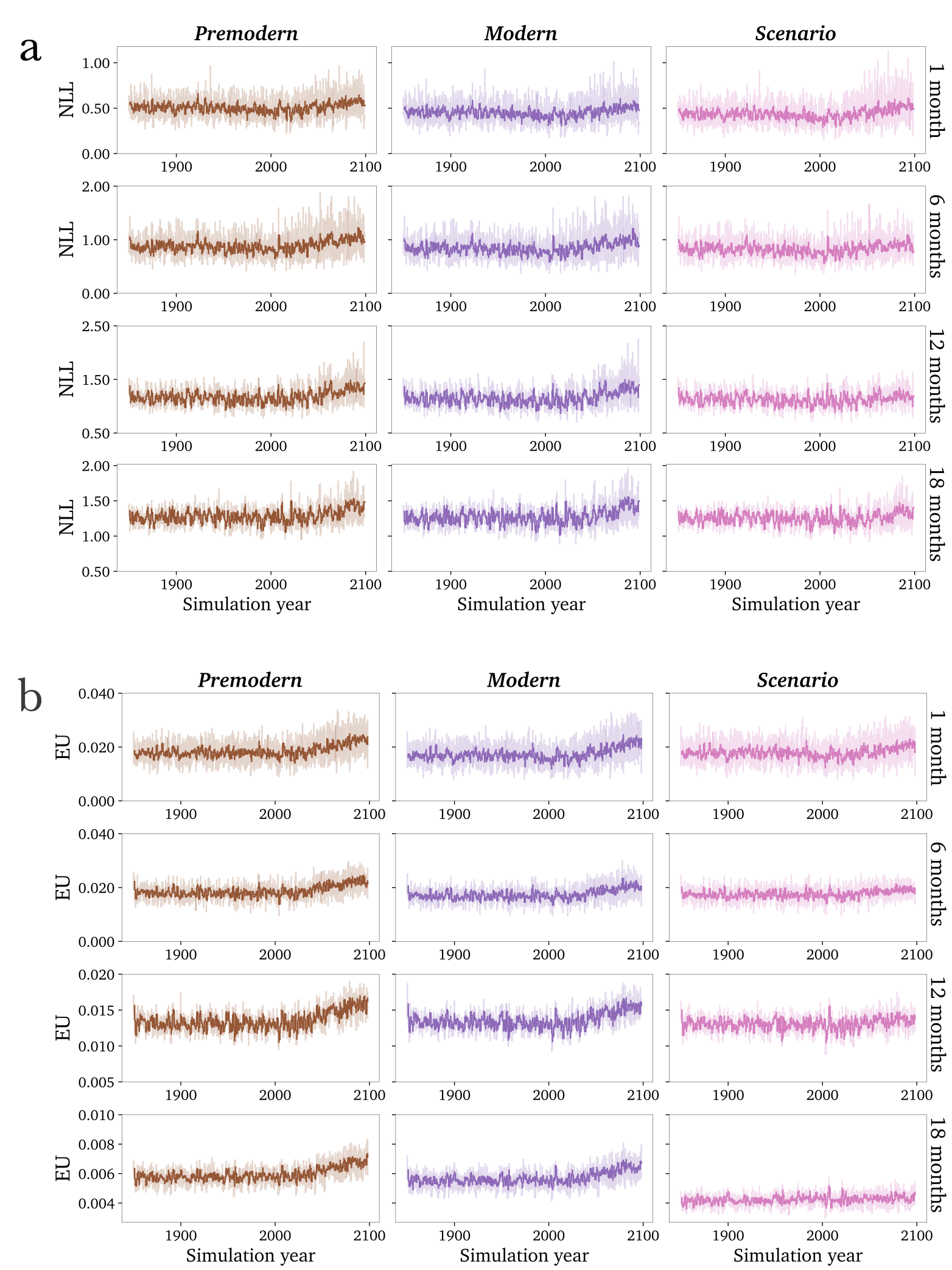}
    \caption{Each ensemble's (a) component-mean NLL and (b) EU averaged over CESM2 testing members; darker lines are 12-month moving averages. The y-axis scaling varies between rows to better highlight temporal trends for each lead. }
    \label{perfshift}
\end{figure}

We now present the key results of this study. Beginning with \fref{perfshift}a, we find that component-mean performance deteriorates starting around the early-to-mid 21st century for each ensemble across lead times. From \fref{perfshift}b, we observe that EU also increases incrementally across leads, in line with NLL, beginning around 2040. This EU increase suggests that covariate shift is the primary mechanism of the observed performance deterioration. As shown in \fref{trends}, EU and NLL increase is distinguishably larger in $\pmodern$ and $\modern$ than in $\scenario$. That is, EU effectively identifies the knowledge deficits of $\pmodern$ and $\modern$ relative to $\scenario$ under future climate change. $\scenario$ is also impacted by the covariate shift, as the shifted period makes up a small fraction of its training data, but is harmed less than $\pmodern$ and $\modern$, for which the projected shift is unfamiliar. 

\begin{figure}[h!]
    \hspace*{-0.5cm} 
    \centering
     \includegraphics[width=0.8\textwidth]{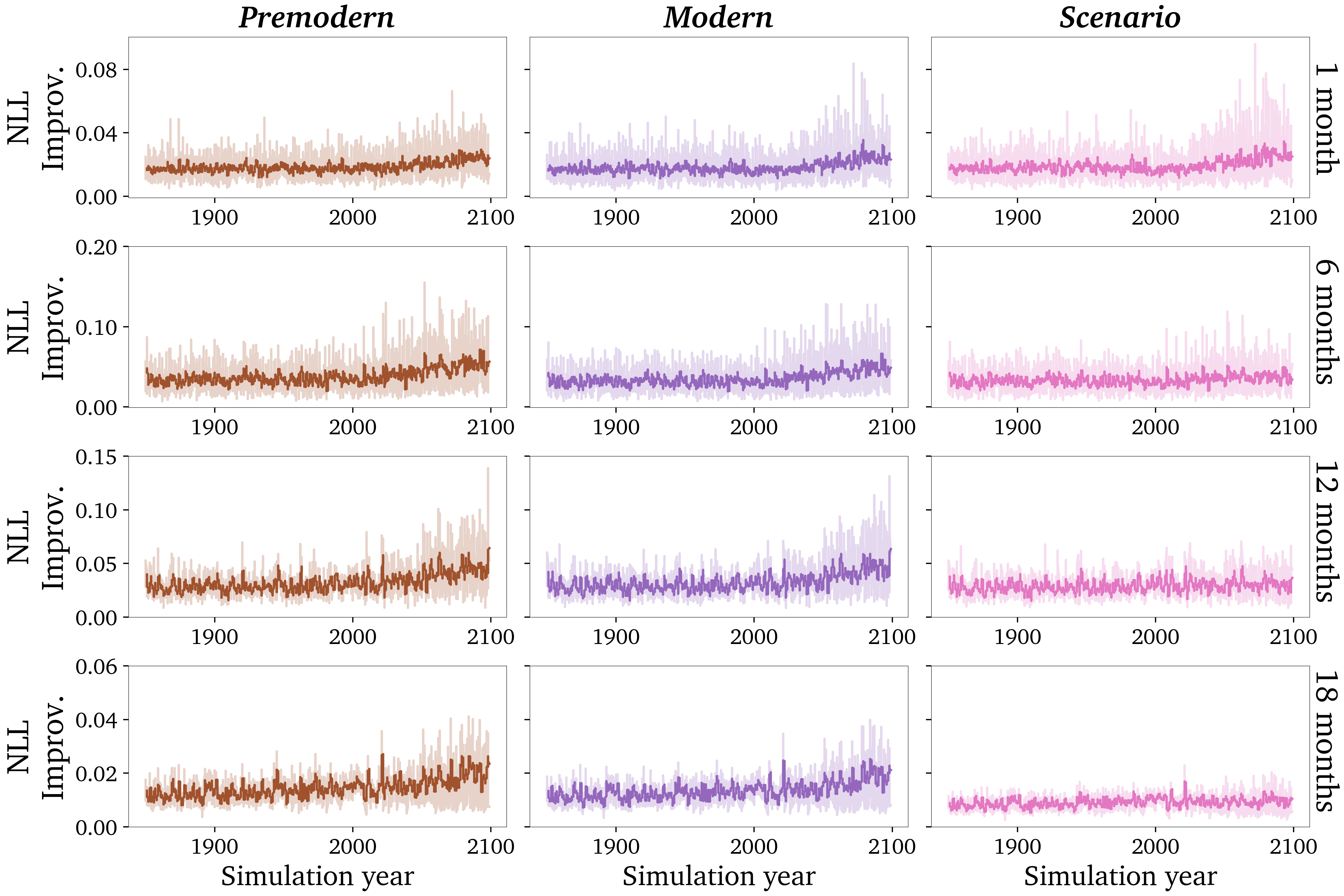}
    \caption{As in \fref{perfshift}, but showing ensemble improvement, i.e., the difference between the component-mean NLL and the ensemble NLL.}
    \label{cov_shift_impr}
\end{figure}

\begin{figure}[h!]
    \hspace*{-0.5cm} 
    \centering
     \includegraphics[width=0.65\textwidth,trim=4 4 4 4, clip]{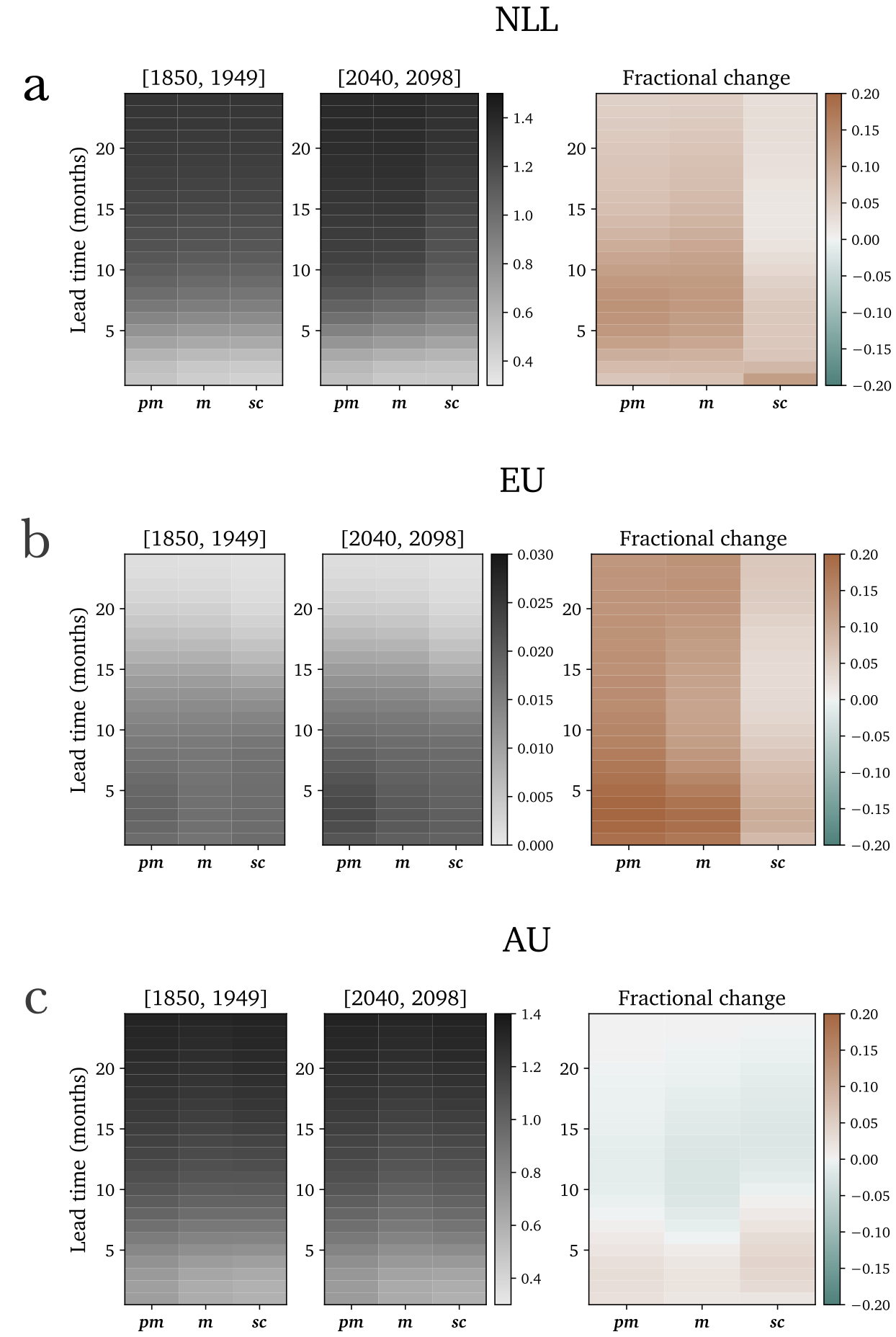}
    \caption{(a) Component-mean NLL, (b) EU, and (c) AU averaged over CESM2 testing members and time, for the premodern and shifted periods, and the fractional increases in each metric from the premodern to shifted periods.  }
    \label{trends}
\end{figure}

Ensemble improvement scales with epistemic uncertainty, as shown in \fref{cov_shift_impr}, because EU measures Brier Score improvement. The incremental growth of NLL improvement indicates that the ensembles deteriorate more slowly under intensifying shift than their components on average. Hence, the utility of deep ensembles to signal covariate shift, via increases in EU, is also what provides protection for ensemble predictions, relative to the expected component, when subject to damaging shift. 

In \fref{trends}c, we plot the change in AU from the premodern period to the shifted period. We find that, counterintuitively, for leads beyond $\sim$ 10 months, mean AU decreases under covariate shift, i.e., component predictions become slightly sharper, even in $\pmodern$ and $\modern$. Fundamentally, this is because AU estimates aleatoric uncertainty, not model confidence, and is thus not designed to recognize performance deterioration under covariate shift. Further, as component performance degrades, AU becomes less reliable at representing aleatoric uncertainty. Consequently, the observed decreases in AU for leads beyond six months under shift do not imply decreased aleatoric uncertainty at those leads -- otherwise performance would improve. Likewise, in \fref{trends}c, the AU increases for leads within $\sim$ 6 months do not necessarily indicate that short-range ENSO predictability decreases under projected climate change, even though short-range performance worsens due to increased epistemic uncertainty.

\begin{figure}[h!]
    \centering
    \includegraphics[width=0.8\linewidth]{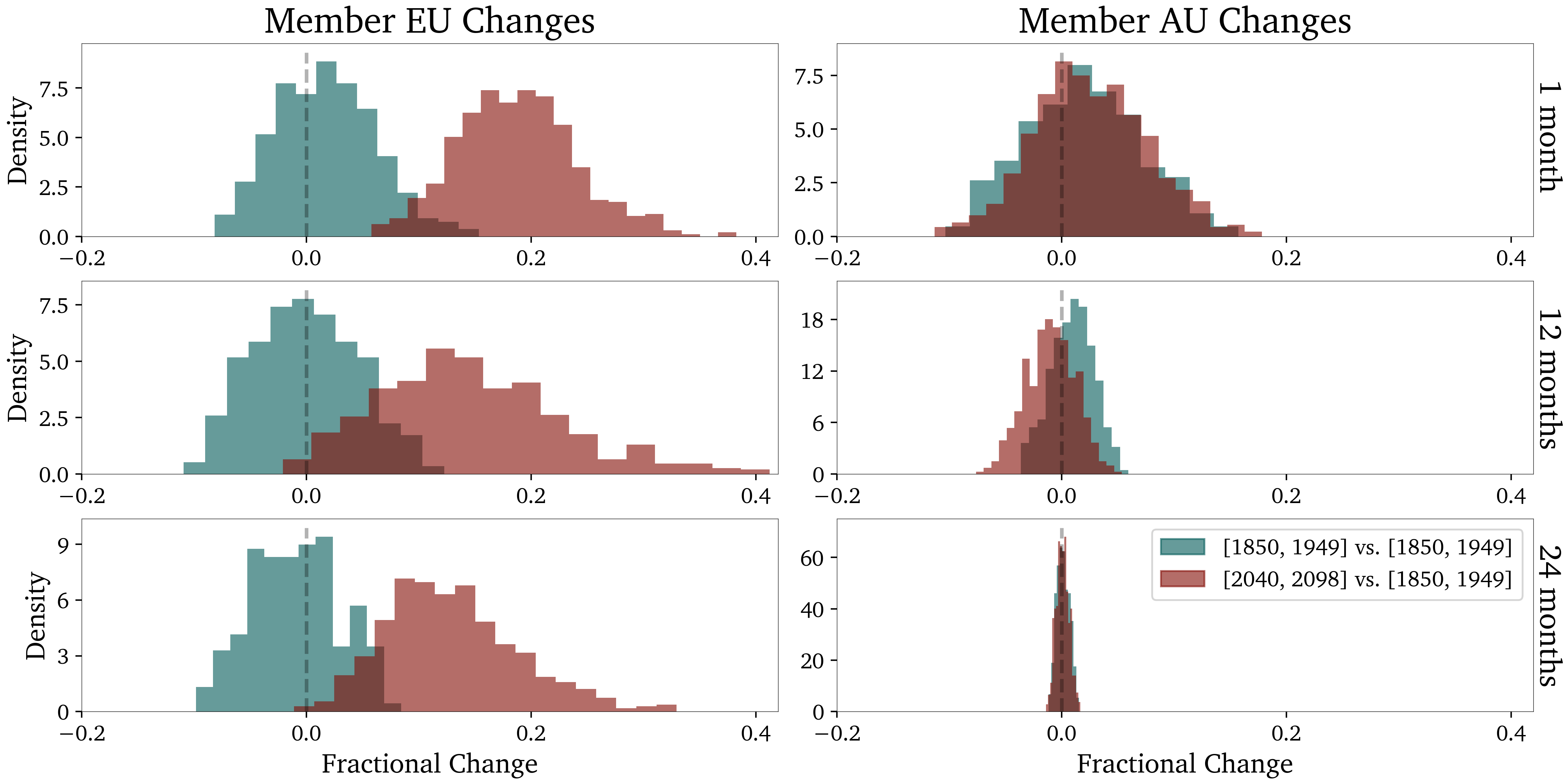}
 
    \caption{Using the $\pmodern$ ensemble predictions, for each CESM2 testing member, we compute the fractional change in AU and EU (relative to the member's own premodern period) when compared against the premodern and shifted periods of every other testing member. Red histograms show comparisons of a testing member's premodern period with the shifted period of another member, while blue histograms show comparisons between premodern periods of different members.}
    \label{change_rel}
\end{figure}

\begin{figure}[h!]
    \centering
    \includegraphics[width=0.85\textwidth]{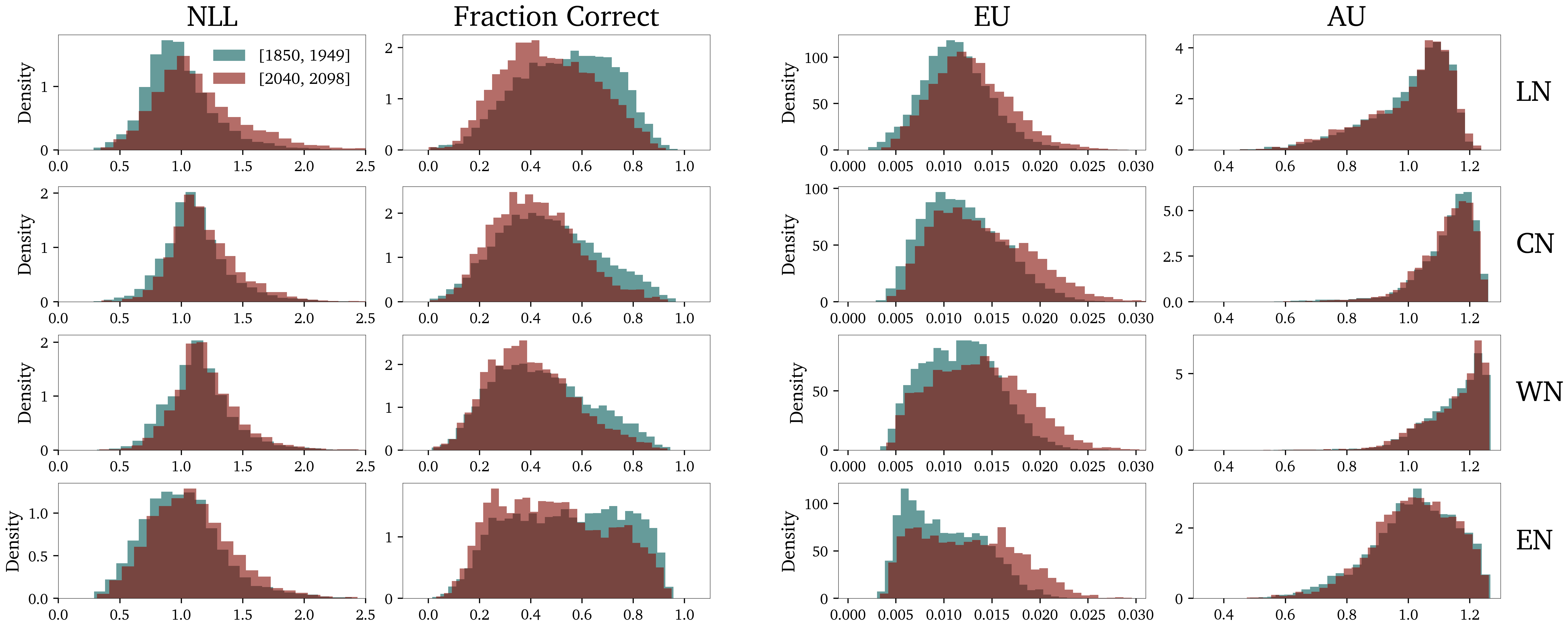}
    \caption{For $\pmodern$, the distribution of component-mean NLL and fraction correct (i.e., the fraction of components which correctly predict the target class), as well as EU and AU over the premodern and shifted periods. For this plot, each metric is averaged over lead times. The mean of the fraction correct distribution is the component-mean accuracy.  }
    \label{cov_shift_phase}
\end{figure}

We further juxtapose the behavior of EU and AU under covariate shift in Figures \ref{change_rel} and \ref{cov_shift_phase}. Notably, as shown in \fref{change_rel}, although internal variability leads to spread in the EU change between CESM2 members, EU consistently increases when comparing samples from a member's premodern period with those from another member's shifted period (as indicated by the positive mean shift of the red EU histograms relative to the blue).  A similar positive AU shift at short leads is present but muted relative to the EU signal, and, unintuitively, the AU change reverses direction at moderate leads. Due to the counterbalancing positive and negative AU changes across lead times, the distribution of lead-time-averaged AU is largely unchanged when comparing the premodern and shifted periods across input ENSO phases, as indicated by the overlapping AU histograms in \fref{cov_shift_phase}. This AU invariance contradicts the NLL and accuracy deterioration observed for each phase, which is further evidence of the decreased reliability of AU under covariate shift. 

Although we observe that, unlike AU, EU robustly signals damaging shift, we emphasize that EU predicts ensemble improvement, not component generalization error itself. For instance, while at each lead time, EU elevations under covariate shift correlate with worsening NLL, \fref{trends} illustrates that the pattern of relative EU increase across leads has dissimilarities with the degradation of NLL. For instance, EU increases at leads $\sim$ 5-10 months correspond to larger relative increases in NLL than the more substantial EU growth seen at shorter leads. Further, as shown in \fref{cov_shift_phase}, increases in the mean and variance of EU under shift are similar across input ENSO phases. Conversely, while performance worsens for all phases, NLL and accuracy deterioration is most notable for LN inputs under shift, as shown in \fref{cov_shift_phase}. Further, forms of distributional shift which have no influence on the input distribution, and are therefore undetectable by EU in the supervised setting \parencite{malininthesis}, can occur in tandem with covariate shift and influence component performance. Hence, EU is a tool for signaling the presence of harmful covariate shift but is not designed for predicting the resultant model errors.

\begin{figure}[h!]
    \centering
     \includegraphics[width=0.95\textwidth]{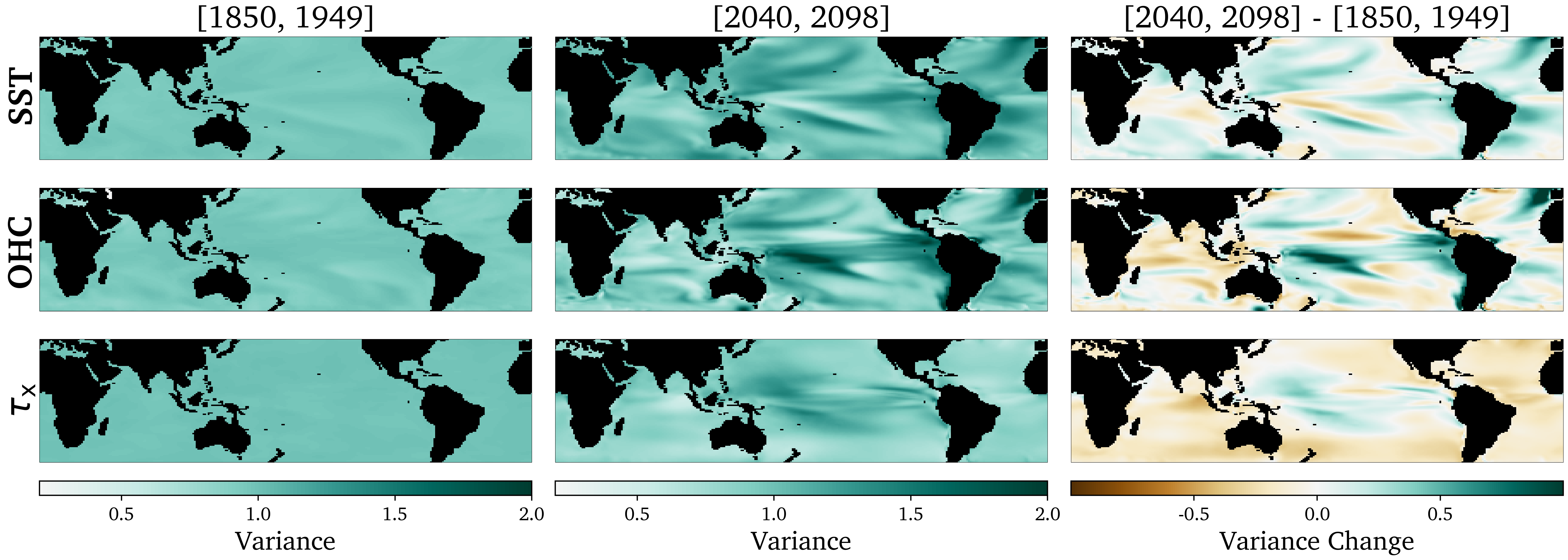}
    \caption{The change in anomaly variance for each input variable from the premodern to shifted period averaged over CESM2 testing members.}
    \label{var_diff}
\end{figure}

The roughly linear EU and NLL increases in \fref{perfshift}, while delayed, resemble the linear forced responses of the global means of SST and OHC under SSP3-7.0 \parencite{ipccch4}. This is despite the fact that the model inputs are anomalies relative to the CESM2-LE mean, and therefore do not contain the forced response. Evidently, the CESM2-LE mean is imperfect at representing the forced response \parencite{frankcombe}, because climate change also influences the variability of the CESM2-LE distribution \parencite{gunnarson, widerange}. Specifically, in \fref{var_diff}, we document substantial changes in the anomaly variance of each input variable under projected climate change. Since the covariate distribution is the joint distribution over input variables, these large, spatially non-uniform changes in marginal variance capture key aspects of the encountered covariate shift. 
 
Although the shifts in extratropical variance in \fref{var_diff} are compelling (we refer to \cite{gunnarson} and \cite{widerange} for relevant discussions), we focus on the tropical Pacific, i.e., the ENSO region, as it has dominant influence on component predictions (see section \ref{sec:xai}). Though we consider different periods and a subset of the CESM2-LE, the SST variance changes in \fref{var_diff} are similar to those identified in \textcite{gunnarson}. The SST variance changes in the tropical Pacific resemble the anomaly signature of El Niño, with decreased variance in the western tropical Pacific, perhaps related to eastward expansion of the Indo-Pacific warm pool \parencite{leung}, and an increase in variance over the eastern tropical Pacific, especially along the equator, perhaps driven by weakening of the Walker Circulation \parencite{held}, as evidenced by increases in equatorial $\tau_x$ variance.  In contrast to SST, the variance of OHC increases substantially over a broad region of the western tropical Pacific, bridged along the equator with another region of increased OHC variance in the eastern tropical Pacific. The spatial patterns of projected variance change in the tropical Pacific are similar for $\tau_x$ and OHC and are potentially linked through ocean-atmosphere coupling. 

\newpage
\section{Discussion}
\label{sec:discussion}

We explore the utility of deep ensembles for ENSO uncertainty quantification. We train ensembles of 100 components -- much larger than what is typical in the literature -- to demonstrate that ensemble performance cannot be replicated by single optimizations of the same architecture. Beyond this, \textcite{wangdisteq} find that ensembling overfit components provides unique benefits to scaling single model complexity, but it is unclear if these powers remain when components are calibrated. 

Well-calibrated components are necessary to ensure in-distribution quality of $\au$ which, in our experiments, properly identifies known characteristics of ENSO predictability, such as the spring predictability barrier and asymmetries in transition uncertainty.  However, AU degrades in quality under covariate shift \parencite{toro}, as predictive uncertainty becomes increasingly epistemic. This is demonstrated in our results, where ensembles exposed to internal variability under an unseen climate change scenario experience a considerable degradation in performance associated with a significant increase in $\eu$. These findings are supported by \textcite{abad}, who train an $M=10$ ensemble of CNNs for downscaling of surface temperature in a global climate model and observe performance deterioration under a future warming scenario, in their case without removing the forced response. Although \textcite{abad} did not report epistemic uncertainty, ensemble improvement, as in our findings, increases under the climate change scenario, indicating that ensemble disagreement scales with intensifying shift. 

Our results support the existing evidence regarding the limitations of single model predictive uncertainty under covariate shift \parencite{galdropout, toro, yangood, hein} in the context of ENSO prediction under future climate change. We find that AU appropriately identifies covariate shift only when increases in EU are substantially large, which occurs mostly for leads within 6 months in our experiments. Contrarily, for longer, less predictable leads, where in-distribution AU is relatively high,  components fail to lower their confidence -- and even become increasingly confident -- at unfamiliar inputs. This is coincident with the observation that, at short lead times, where mean AU is low, there is a moderate correlation between AU and EU -- consistent with, e.g., the $M=20$ ensemble of \cite{schreck} in a low AU hydrometeor classification problem -- while this correlation breaks down and reverses as AU increases with progressing lead (see \fref{au_eu_corr}). Many relevant problems in climate have high aleatoric uncertainty. Our results suggests that, in such scenarios, not accounting for epistemic uncertainty under covariate shift can be particularly costly.

Aside from variational inference, further exploration into cheap single model alternatives to ensembles for joint aleatoric and epistemic UQ is warranted. For instance, evidential deep learning (EDL) \parencite{sensoy}, also known as Prior Networks \parencite{malinin}, tasks a neural network to model a Dirichlet distribution over predictive categoricals, where the spread of the Dirichlet captures epistemic uncertainty. EDL performs well on benchmark datasets and shows promise in \textcite{schreck} for meteorological tasks. However, the family of loss functions used in these techniques can lead to undesirable properties of the learned epistemic uncertainty, like an insensitivity to training size \parencite{bengs, shen}. A related approach is ensemble distribution distillation \parencite{malinin, ryabanin}, which attempts to distill the epistemic distributions of a deep ensemble in a single model to save time on inference computation. This method has shown performance improvements over traditional knowledge distillation but struggles to capture increases in epistemic uncertainty away from the training distribution \parencite{malinin}. Beyond deep ensembles, the investigation of more theoretically grounded frameworks for deep UQ, e.g., Bayesian deep ensemble methods \parencite{dangelo, wildlink}, is a promising avenue for future research at the intersection of ML and climate. 

\section*{Acknowledgements}
Funding for this project was provided in part by grants AGS-2210068 and AGS-1749261 from the National Science Foundation. We thank John Fasullo at the National Center for Atmospheric Research for providing the OHC data. 

\section*{Open Research}
Code for this project is available at \url{https://github.com/dev-mcafee/deepensembles}. 

\setcounter{figure}{0}

\printbibliography   

\ifincludesupp
    \begin{refsection}
\newif\ifincludebib   
\includebibfalse
\documentclass[Revised_Manuscript.tex]{subfiles}
\externaldocument{Revised_Manuscript}
\usepackage[
    maxcitenames=2,
    maxbibnames=3,
    style=authoryear,
    backend=biber
]{biblatex}

\setlength{\parskip}{0.5em} 

\usepackage[a4paper, left=1in, right=1in, top=1in, bottom=1in]{geometry}

\title{\Large \textbf{Am I Confused or Is This Confusing?: \\  Deep Ensembles for ENSO Uncertainty Quantification} \\ \textit{Supplemental Material}}

\begin{document}
\maketitle
\appendix
\renewcommand{\appendixname}{Supplemental Material}
\renewcommand{\appendixpagename}{Supplemental Material}
\renewcommand{\thesection}{S\arabic{section}}
\renewcommand{\thefigure}{S\arabic{figure}}
\renewcommand{\thetable}{S\arabic{table}}
\renewcommand{\theequation}{S\arabic{equation}}

\section{Regularization}
\label{sec:prior}

The addition of a prior $p(\wvec)$ to the NLL objective makes the training loss the negative log posterior density. 

\begin{equation*}
    L(\pmodel) =  -\left(\sum_{j=1}^{N}\log{\pmodel(y_j \mid \xvec_j)} + \log{p(\wvec)} \right)
\end{equation*}

\begin{equation*}
    =  -\left( \log{\prod_{j=1}^{N}\pmodel(y_j \mid \xvec_j)}+  \log{p(\wvec)} \right)
\end{equation*}

\begin{equation*}
    \stackrel{c}{=} - (\log \pmodel(\data) + \log{p(\wvec)})
\end{equation*}

\begin{equation*}
    \stackrel{c}{=} - \log p(\wvec \mid \data)
\end{equation*}

where $\stackrel{c}{=}$ denotes equality up to an additive constant. The standard approach in Bayesian deep learning assumes a zero-mean, isotropic Gaussian prior; in \textcite{izmailovhmc}, the isotropic Gaussian yields similar performance to a logistic prior and a mixture of two isotropic Gaussians. This is equivalent to L2 regularization. 

\begin{equation*}
\log p(\wvec) = \log \prod_{l=1}^{N_w} 
    \frac{1}{\sqrt{2 \pi \sigma^2}} 
    \exp\left( - \frac{(\wvec^{(l)})^2}{2\sigma^2} \right)
\end{equation*}

\begin{equation*}
    = \sum_{l=1}^{N_w} \log\left(
        \frac{1}{\sqrt{2 \pi \sigma^2}} 
        \exp\left( - \frac{(\wvec^{(l)})^2}{2\sigma^2} \right)\right)
    \end{equation*}

\begin{equation*}
    \stackrel{c}{=} - \frac{1}{2\sigma^2}\|\mathbf{\wvec}\|_2^2
\end{equation*}

where $N_w$ is the number of weights and $\sigma^2$ is the prior variance.

\section{EU and Ensemble Improvement}
\label{sec:brierimprov}

This derivation is also provided in, e.g.,  \textcite{abePredictiveDiversity}. 

Let $\textbf{p}_i$ denote the probability vector predicted by component $i$ of a deep ensemble, given an arbitrary input with target $y$, and $\bar{\textbf{p}}$ represent the parameters of the ensemble-mean predictive distribution. Brier Score improvement (over the component-mean score) is guaranteed by Jensen's inequality. 

\begin{equation*}
    \frac{1}{M}\sum_{i=1}^{M}\text{BS}(\textbf{p}_i) - \text{BS}(\bar{\textbf{p}}) = \frac{1}{M}\sum_{i=1}^{M}\sum_{k=1}^{K}(\textbf{p}_i^{(k)} - [y=k])^2 - (\bar{\textbf{p}}^{(k)} - [y=k])^2
\end{equation*}

\begin{equation*}
    = \frac{1}{M}\sum_{i=1}^{M}\sum_{k=1}^{K}(\textbf{p}_i^{(k)})^2 - 2[y=k] \textbf{p}_i^{(k)} + [y=k]^2 - (\bar{\textbf{p}}^{(k)})^2 + 2[y=k]\bar{\textbf{p}}^{(k)} - [y=k]^2
\end{equation*}

\begin{equation*}
    = \frac{1}{M}\sum_{k=1}^{K}\sum_{i=1}^{M}(\textbf{p}_i^{(k)})^2 - (\bar{\textbf{p}}^{(k)})^2  + \frac{2[y=k]}{M}\sum_{k=1}^{K}\sum_{i=1}^{M}(\bar{\textbf{p}}^{(k)} - \textbf{p}_i^{(k)})
\end{equation*}

\begin{equation*}
    = \frac{1}{M}\sum_{k=1}^{K}\sum_{i=1}^{M}(\textbf{p}_i^{(k)})^2 - (\bar{\textbf{p}}^{(k)})^2
\end{equation*}

This is equivalent to equation \eqref{eu_eq}.

For convex loss $\phi$, $\phi(\text{E}[p], y) - \text{E}[\phi(p, y)]$ is generally referred to as the Jensen gap, i.e., Brier Score improvement is the Jensen gap with respect to Brier Score. Beyond Brier Score, we expect the Jensen gap of any convex loss to increase under covariate shift if the variance of ensemble probabilities (EU) increases \parencite{liaosharp, streeter}. For example, in \fref{cov_shift_impr}, the Jensen gap with respect to NLL increases with EU under covariate shift.

\section{Fixed Lead Design}

\label{sec:fltd}
Below, we reproduce the figures of section \ref{sec:results} using the fixed lead time (flt) problem design, where we train independent ($M=10$) ensembles for leads 1, 6, 12, and 18 months. 

\begin{figure}[h!]
    \hspace{-0.5cm}
    \centering
     \includegraphics[width=0.9\textwidth,trim=4 4 4 4, clip]{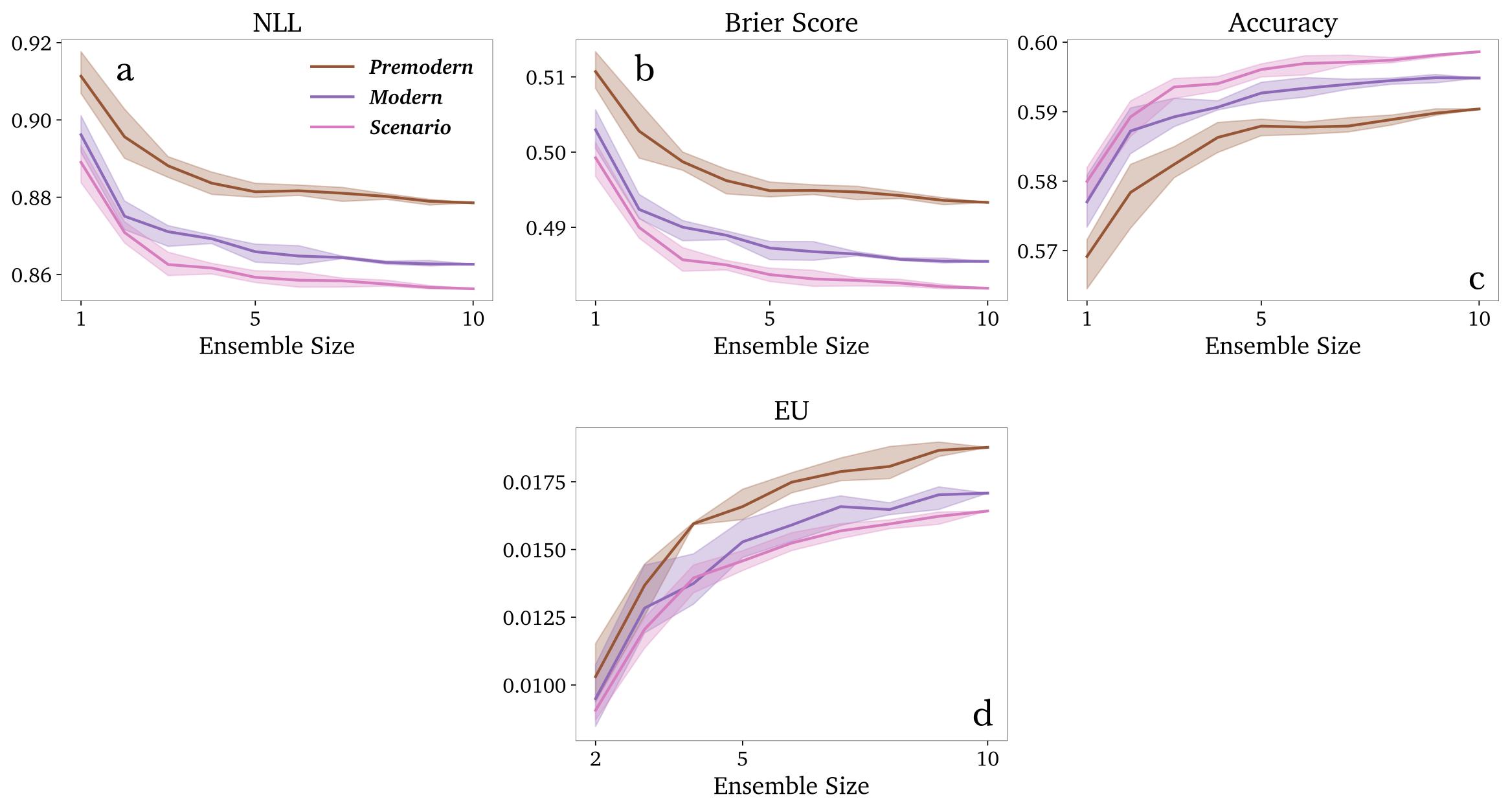}
     \caption{As in \fref{capacity} but for flt design. Increasing ensemble size improves performance of the ensemble-mean predictive distribution by better accounting for epistemic uncertainty.}
    \label{capacity_flt}
\end{figure}

\begin{figure}[h!]
    \centering
     \includegraphics[width=0.9\textwidth]{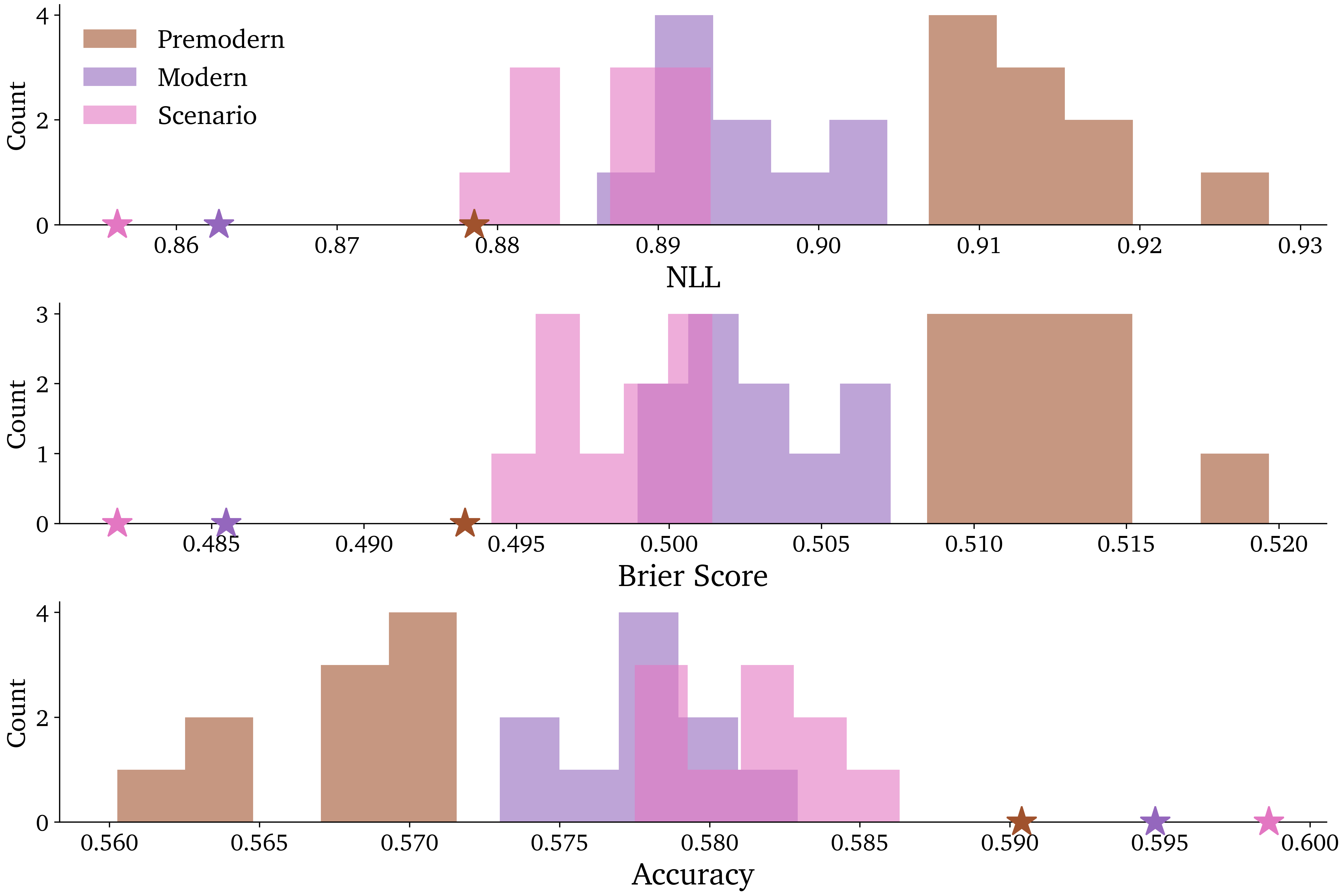}
    \label{improvement_flt}
    \caption{As in \fref{improvement} for flt design. When averaging skill across lead times, each ensemble outscores all of its components.}
\end{figure}

\begin{figure}[h!]
    \centering
     \includegraphics[width=0.8\textwidth]{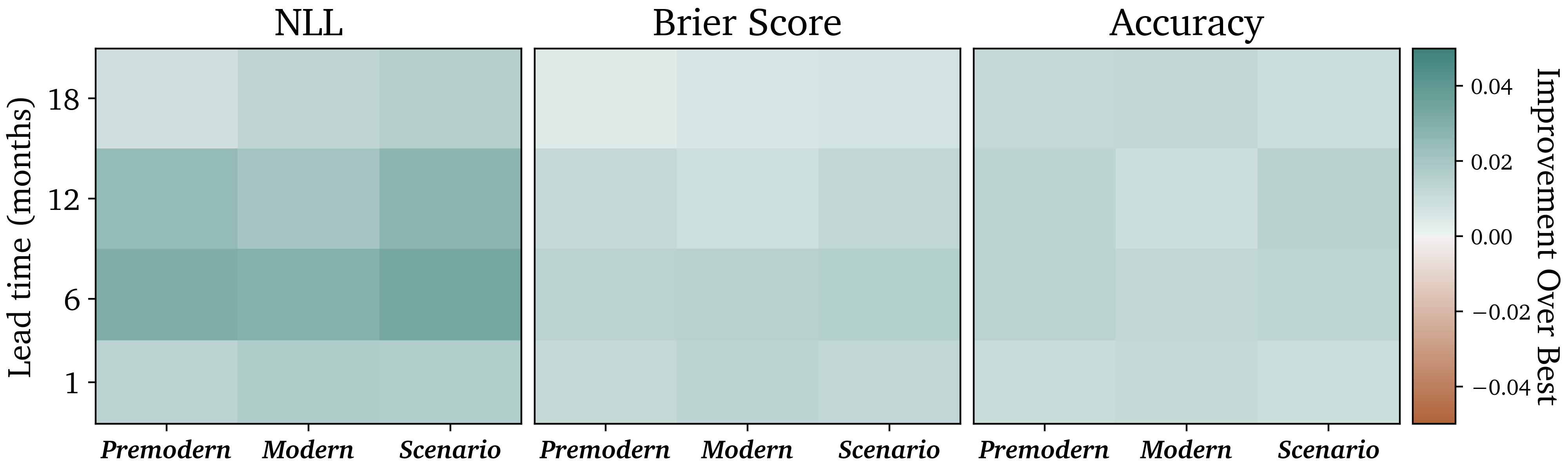}
    \caption{As in \fref{ens_vs_best} for flt design. Each ensemble outperforms all of its individual components for each lead time. }
    \label{ens_vs_best_flt}
\end{figure}

\begin{figure}
    \centering
     \includegraphics[width=0.9\textwidth]{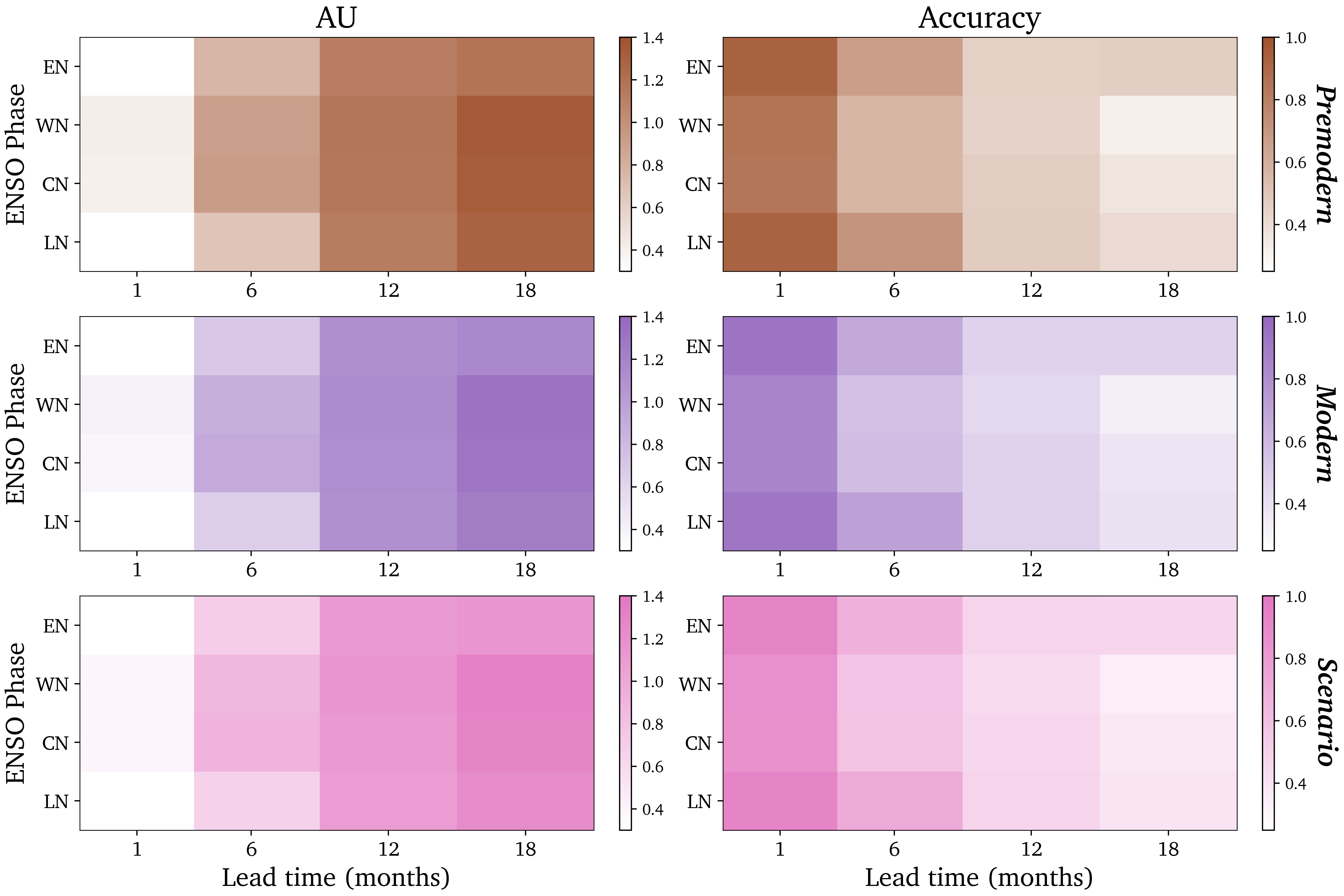}
     \caption{As in \fref{unc_phase} for flt design. Predictability is lower for neutral phases and 18 month predictability is higher for EN than LN initializations.}
    \label{unc_phase_flt}
\end{figure}

\begin{figure}
    \centering
     \includegraphics[width=0.9\textwidth]{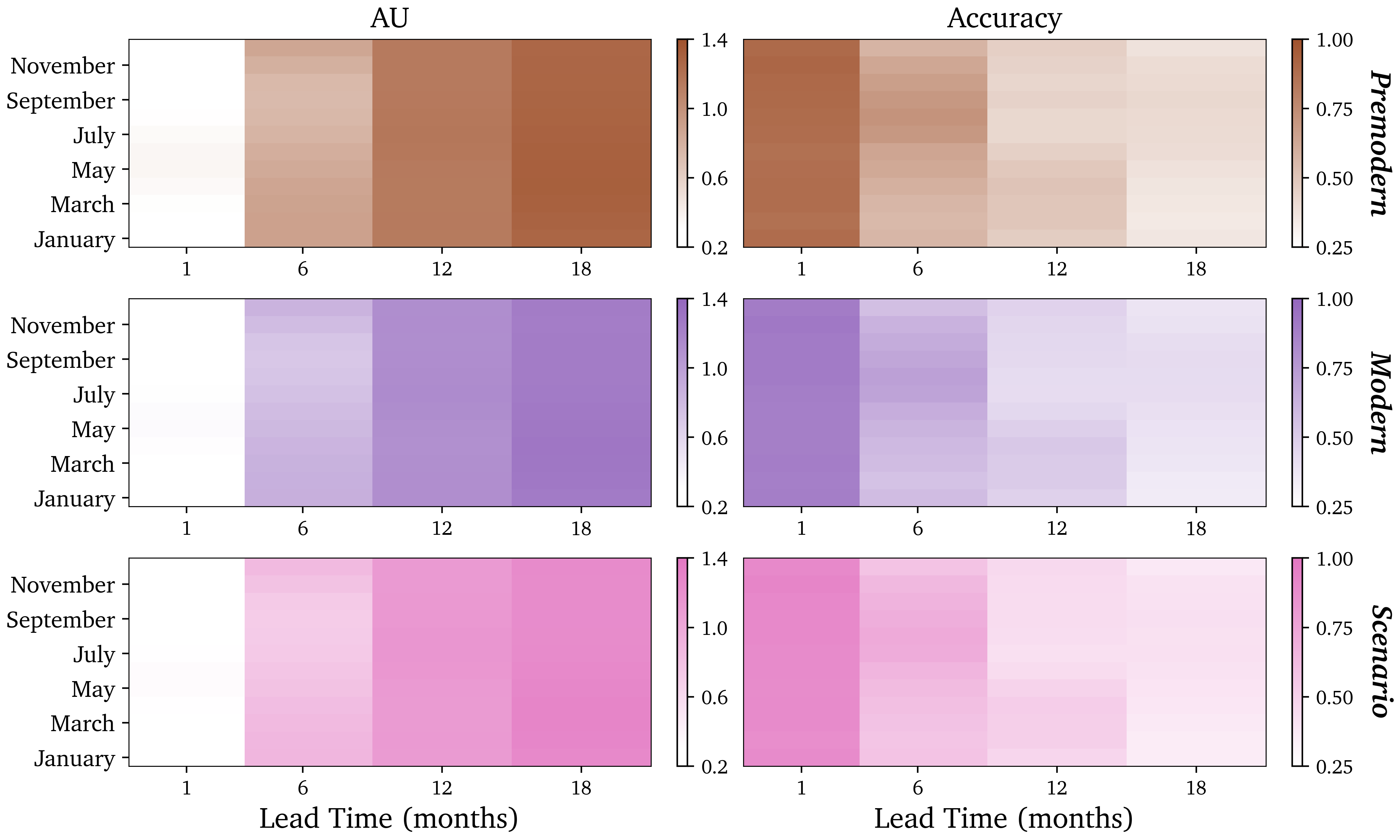}
     \caption{As in \fref{seasonality} for flt design. AU identifies winter phase locking and spring predictability barrier. }
    \label{seasonality_flt}
\end{figure}

\begin{figure}[h!]

    \centering
     \includegraphics[width=0.8\textwidth,trim=4 4 4 4, clip]{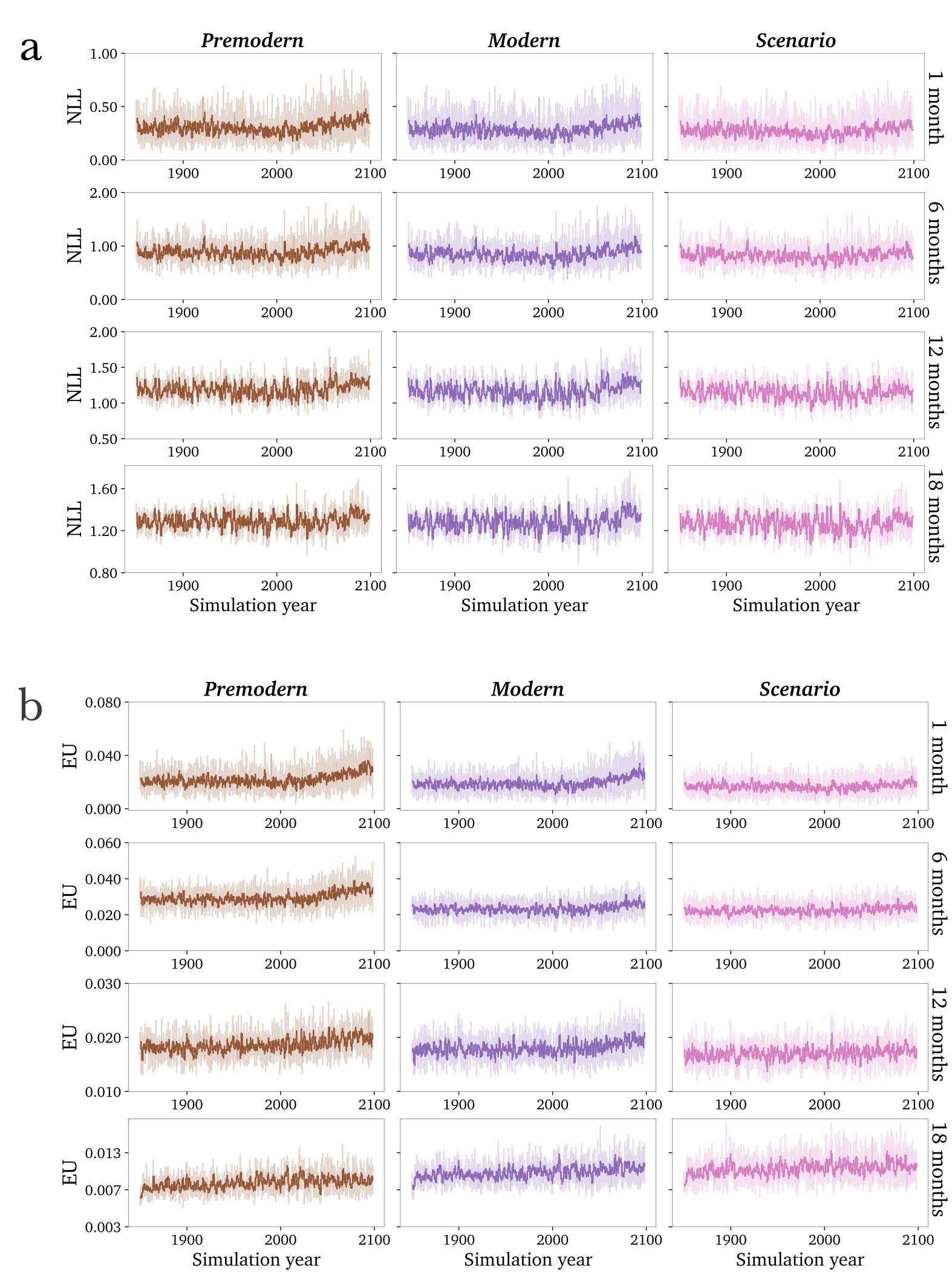}
    \caption{As in \fref{perfshift} for flt design. EU and component-mean NLL increase after the mid-21st century.}
    \label{perfshift_flt}
\end{figure}

\begin{figure}[h!]
    \hspace*{-0.5cm} 
    \centering
     \includegraphics[width=0.8\textwidth]{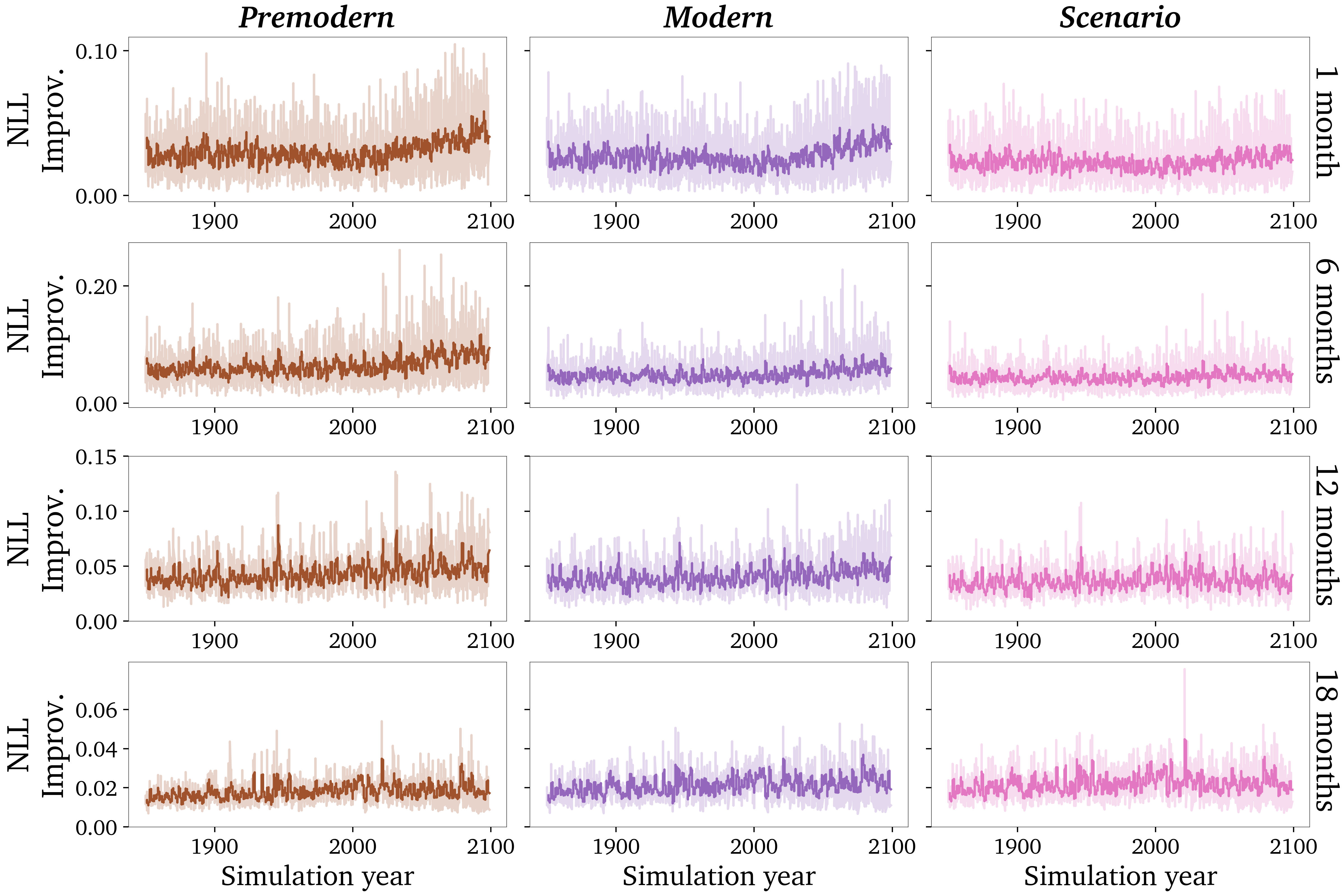}
    \caption{As in \fref{cov_shift_impr} for flt design. NLL improvement increases after the mid-21st century in line with EU.}
    \label{cov_shift_impr_flt}
\end{figure}

\begin{figure}[h!]
    \hspace*{-0.5cm} 
    \centering
     \includegraphics[width=0.7\textwidth,trim=4 4 4 4, clip]{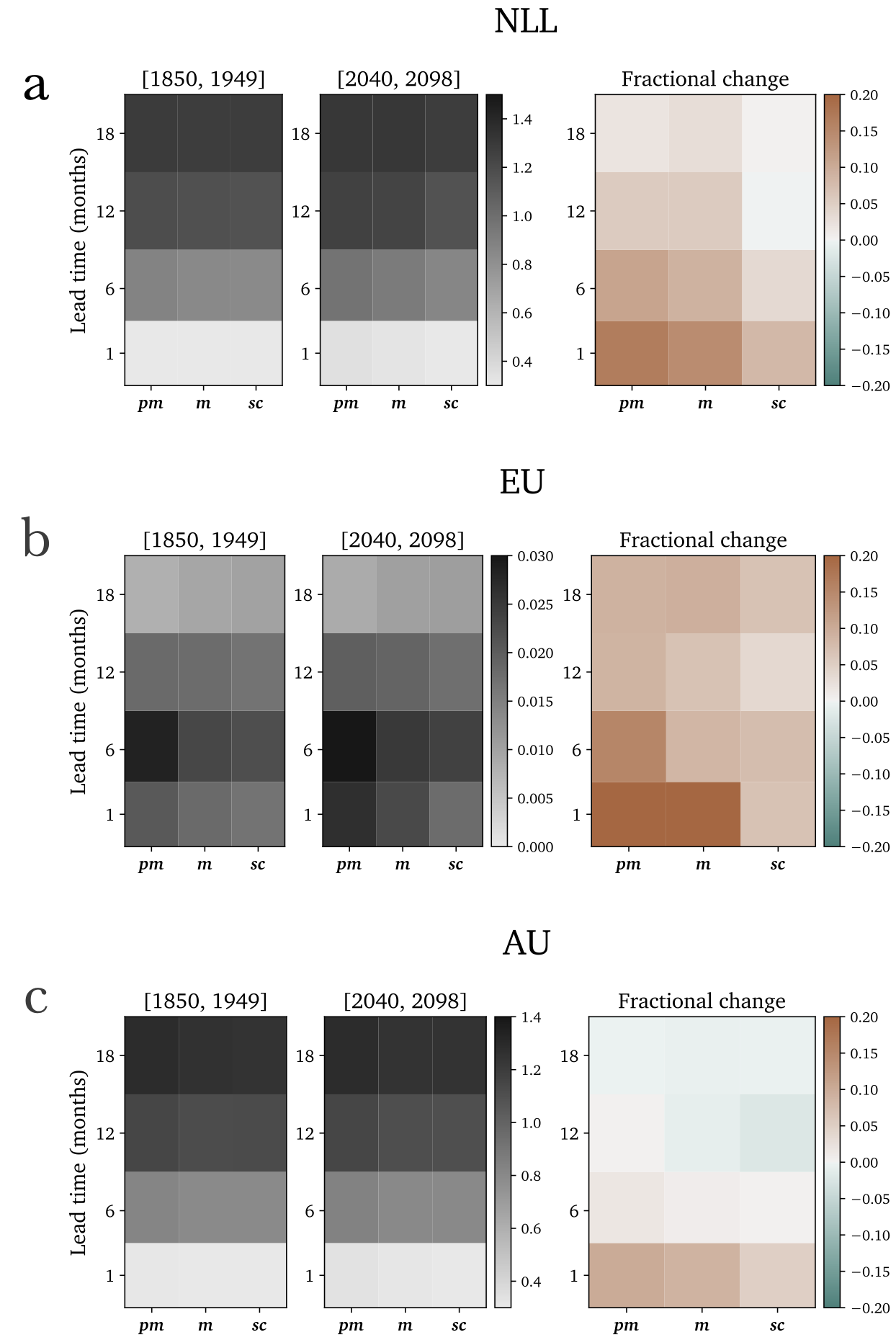}
    \caption{As in \fref{trends} for flt design. AU increases under covariate shift only when EU increase is sufficiently large.}
    \label{trends_flt}
\end{figure}

\begin{figure}[h!]
    \centering
    \includegraphics[width=0.83\linewidth]{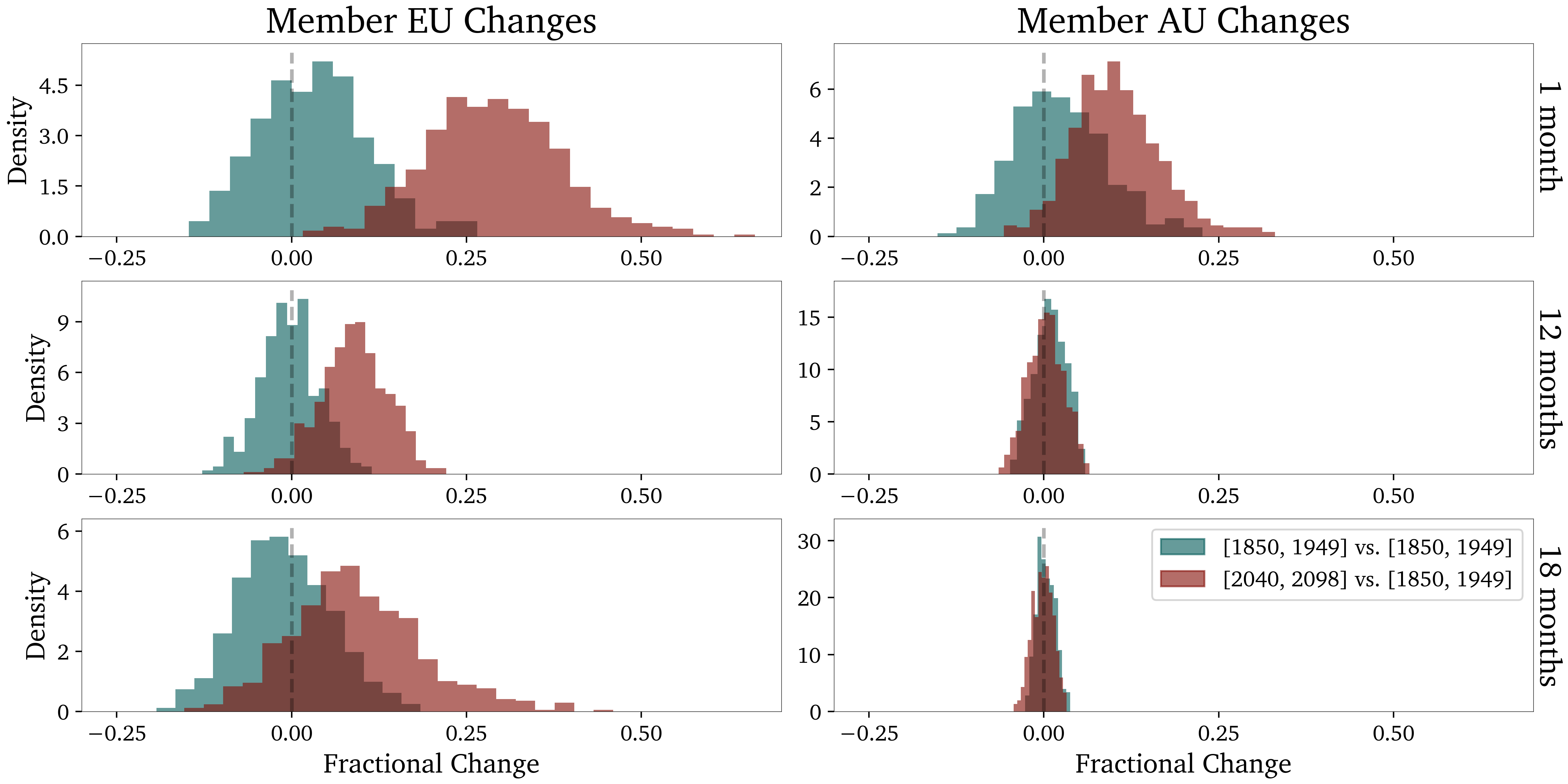}

    \caption{As in \fref{change_rel} for flt design for leads 1, 12 and 18 months. EU better differentiates member samples from the premodern period versus the shifted period than AU.}

    \label{change_rel_flt}
\end{figure}  

\begin{figure}[h!]
    \centering
     \includegraphics[width=0.9\textwidth]{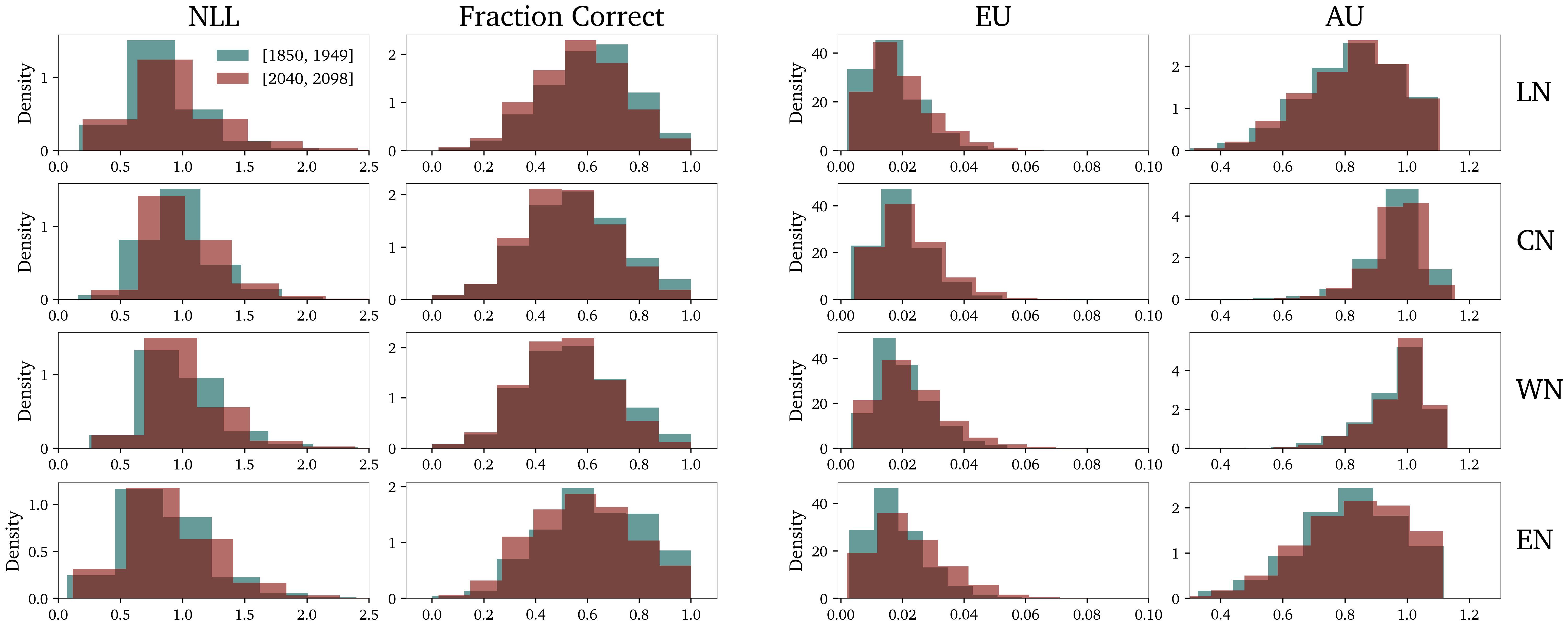}
    \caption{As in \fref{cov_shift_phase} for flt design. Loss and EU increase for each phase, while the AU signal is dampened.}
    \label{cov_shift_phase_flt}
\end{figure}

\clearpage
\section{Calibration Bias} 
\label{sec:clb}

\begin{figure}[h]
    \centering
     \includegraphics[width=0.9\textwidth]{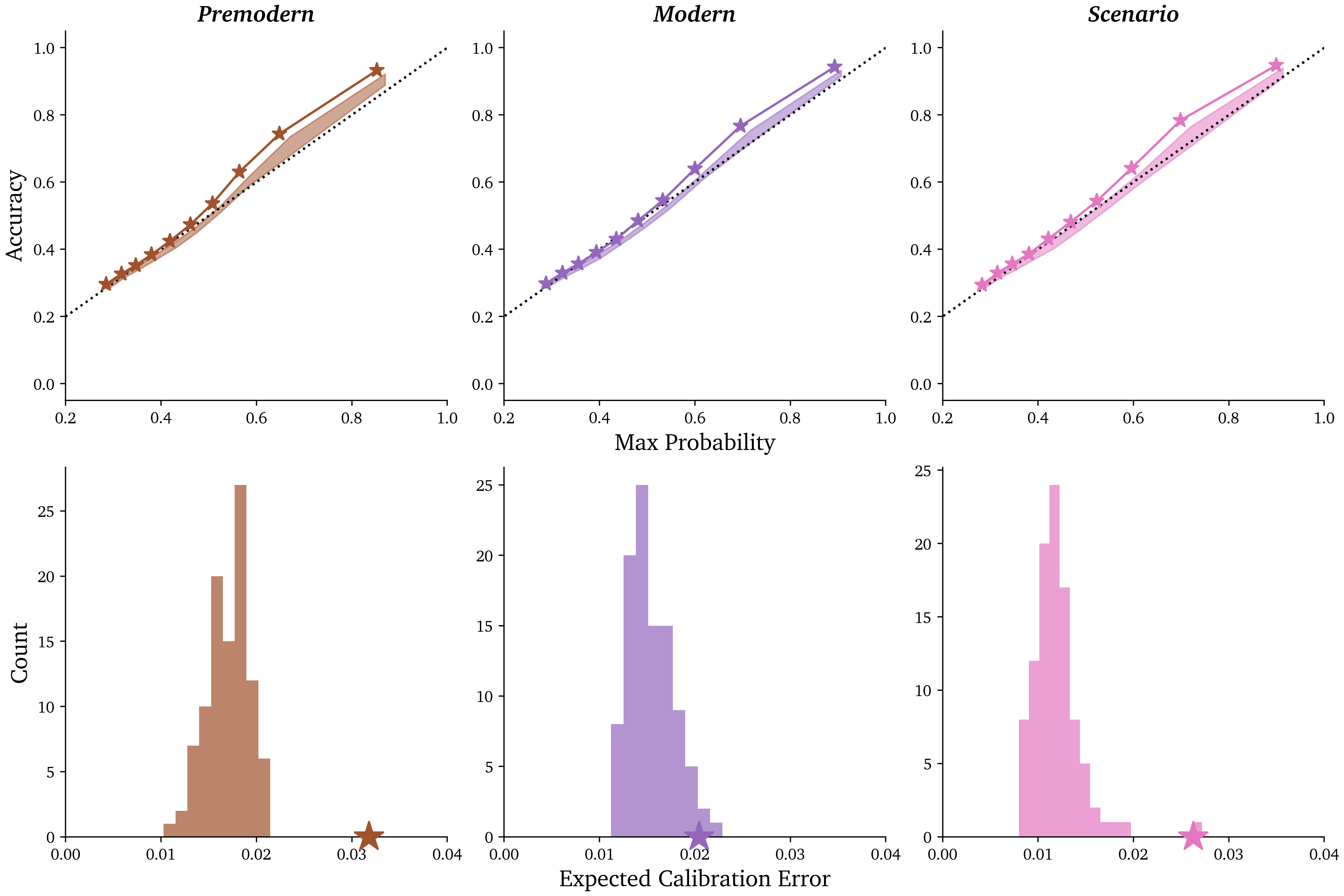}
     \caption{(a) Top-label reliability diagrams for each ensemble (stars) and their components (shading) and (b) corresponding expected calibration errors for the premodern period.}
    \label{reliability}
\end{figure}

\begin{figure}[h]
    \centering
     \includegraphics[width=0.9\textwidth]{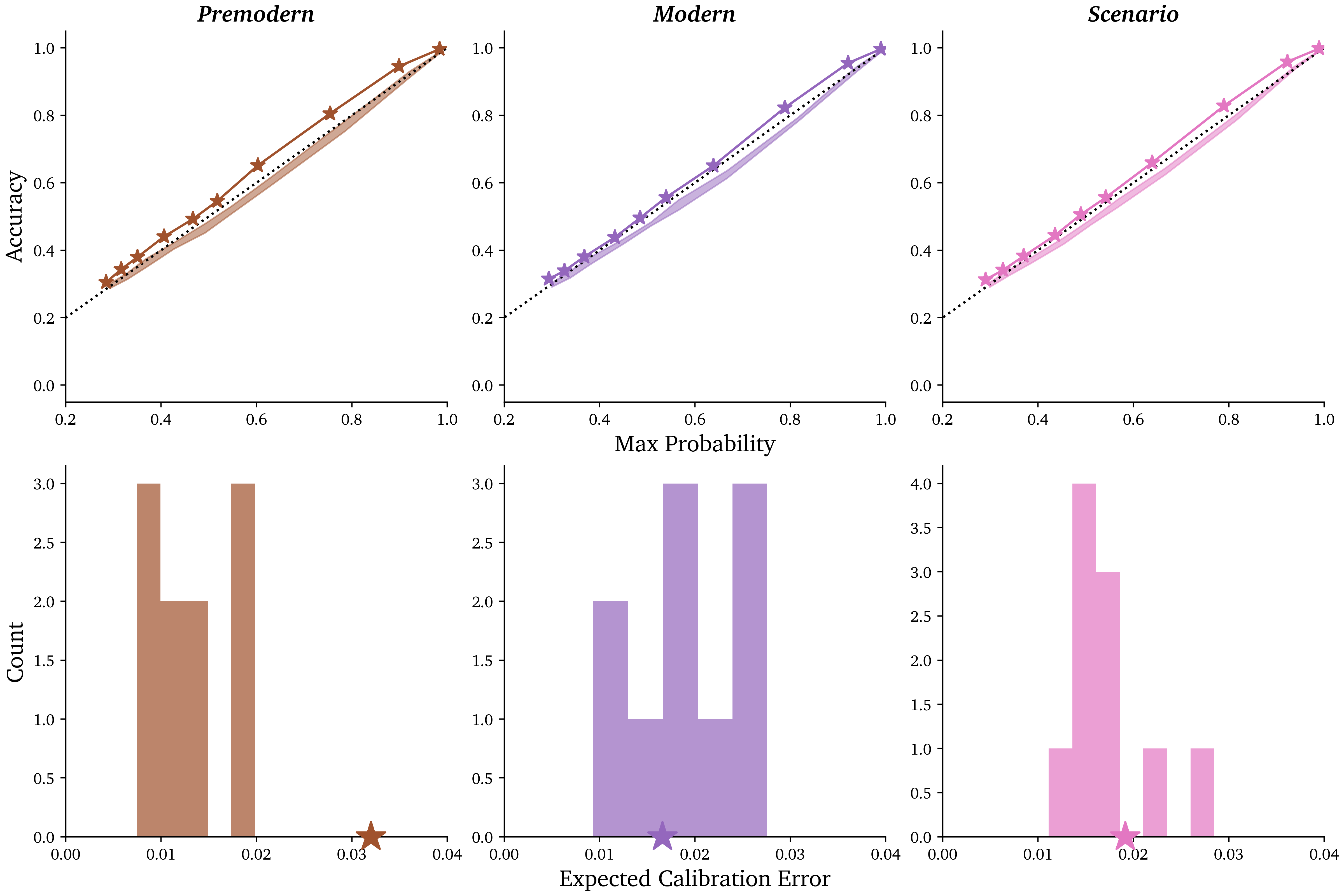}
     \caption{As in \fref{reliability} for flt design. }
    \label{reliability_flt}
\end{figure}

In \fref{reliability}, we show the top-label (predicted class) calibration of each ensemble and its components \parencite{guo}. We group lead times together in this calibration analysis for ease of visualization. Let $y_0$ represent the prediction of the classifier induced by model $p$. The model is top-label calibrated if it satisfies:

\begin{equation*}
    \mathbb{E} [[y=y_0] \mid p(y_0)] = p(y_0)
\end{equation*}

Top-label calibration requires that, e.g., across all inputs where a model predicts an arbitrary ENSO phase with 60\% probability, that phase transpires roughly 60\% of the time. Top-label calibration is visualized using reliability diagrams \parencite{murphyreliability}, where a model's top-label probabilities over the testing set are grouped into, e.g., equally-populated bins \parencite{naeini}, and the accuracy of each bin is plotted against the bin-mean probabilities. The reliability curve of a top-label calibrated model will approximately follow the 1:1 dashed line of the reliability diagram \parencite{brocker}. The expected calibration error (ECE), or the absolute deviation from the 1:1 line averaged across bins, is a commonly reported metric of top-label calibration and approaches zero at calibration \parencite{naeini}.

\fref{reliability} shows that, across ensembles, the components have good top-label calibration, and the component-mean ECE improves marginally with training size. However, surprisingly, we find in \fref{reliability}b that the ensembles, while still well calibrated, have worse ECE than almost all of their components. This conflicts with the consensus in the literature that deep ensembling improves calibration, as measured by ECE \parencite{ovadia}. From \fref{reliability}a, we observe an underestimation bias in top-label ensemble probabilities, predominantly affecting short-range predictions. Although less pronounced, the components also demonstrate slight underestimation bias at short leads, which is likely due to training interference from longer leads. However, \fref{reliability_flt} shows that this systematic bias in ensemble calibration remains when using the fixed lead strategy.  

Deep ensemble underestimation bias is also observed in \textcite{rahaman} and \textcite{wu}, who provide similar explanation for this phenomenon. Informally, this artifact is a consequence of the simultaneous increases in both predictive accuracy and uncertainty when averaging probabilistic predictions. Namely, the entropy of the ensemble prediction is at least the component-mean entropy, i.e., AU, according to Jensen's inequality \parencite{rahaman}. Likewise, the ensemble's top-label probability can be no greater than the component-mean top-label probability \parencite{wu}. Thus, while ensemble averaging generally improves accuracy within bins, it simultaneously prohibits the bin-centers from migrating to higher probabilities relative to less-accurate components. This deteriorates ECE unless ensemble components have overestimation bias, which is common in deep learning \parencite{guo, ovadia}, perhaps related to overfitting on insufficient training data \parencite{blasiokproper}. Importantly, this artifact is not specific to deep ensembles and can affect all methods which improve accuracy by averaging over probabilistic predictions, such as VI or Bayesian ensemble methods. 

\begin{figure}
    \centering
    \includegraphics[width=0.6\textwidth]{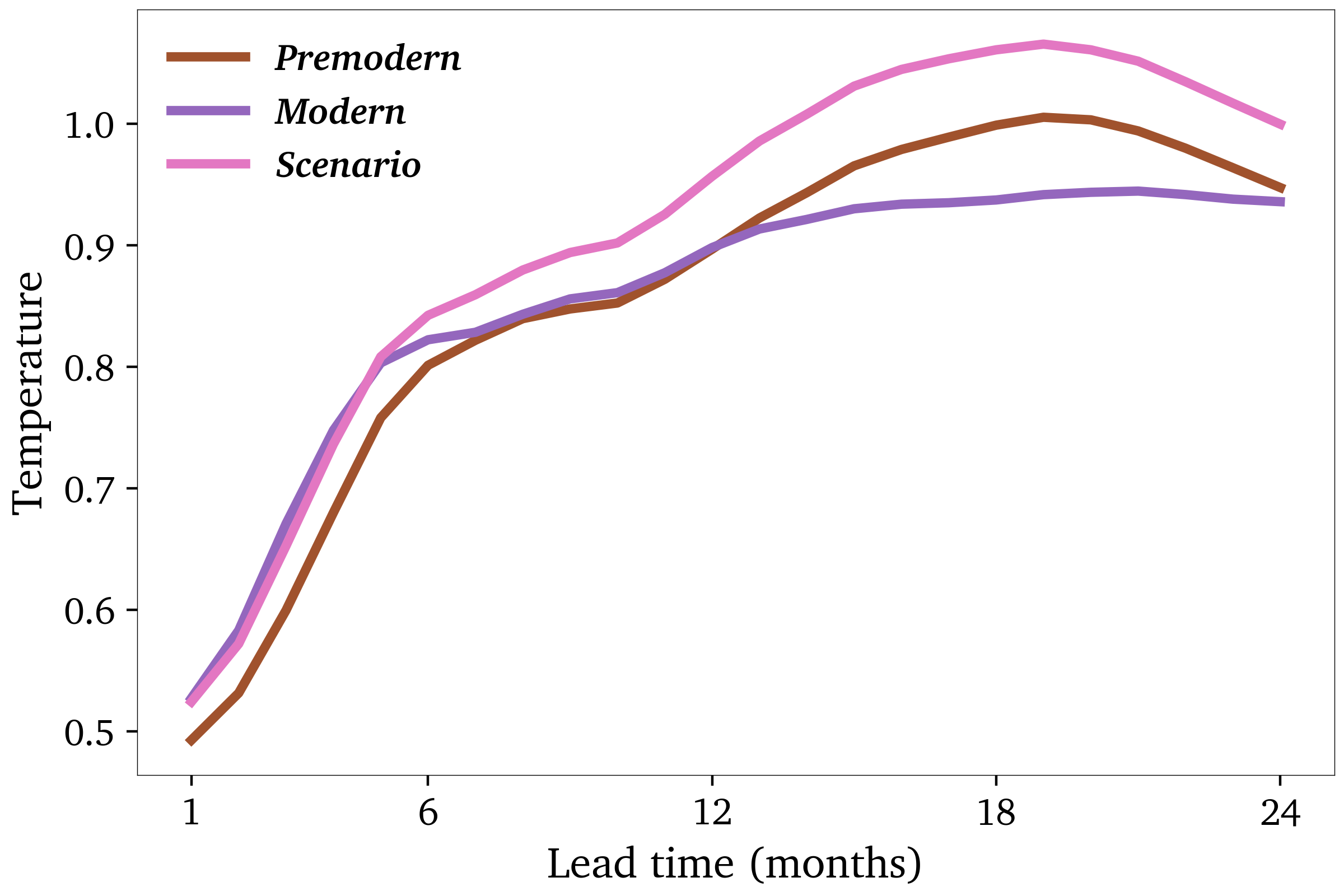}
    \caption{Scaled temperatures for each ensemble. Due to component biases, ensemble underestimation bias is largest for leads within 5 months. }
\label{temps}
\end{figure}

\begin{figure}
    \centering
     \includegraphics[width=0.9\textwidth]{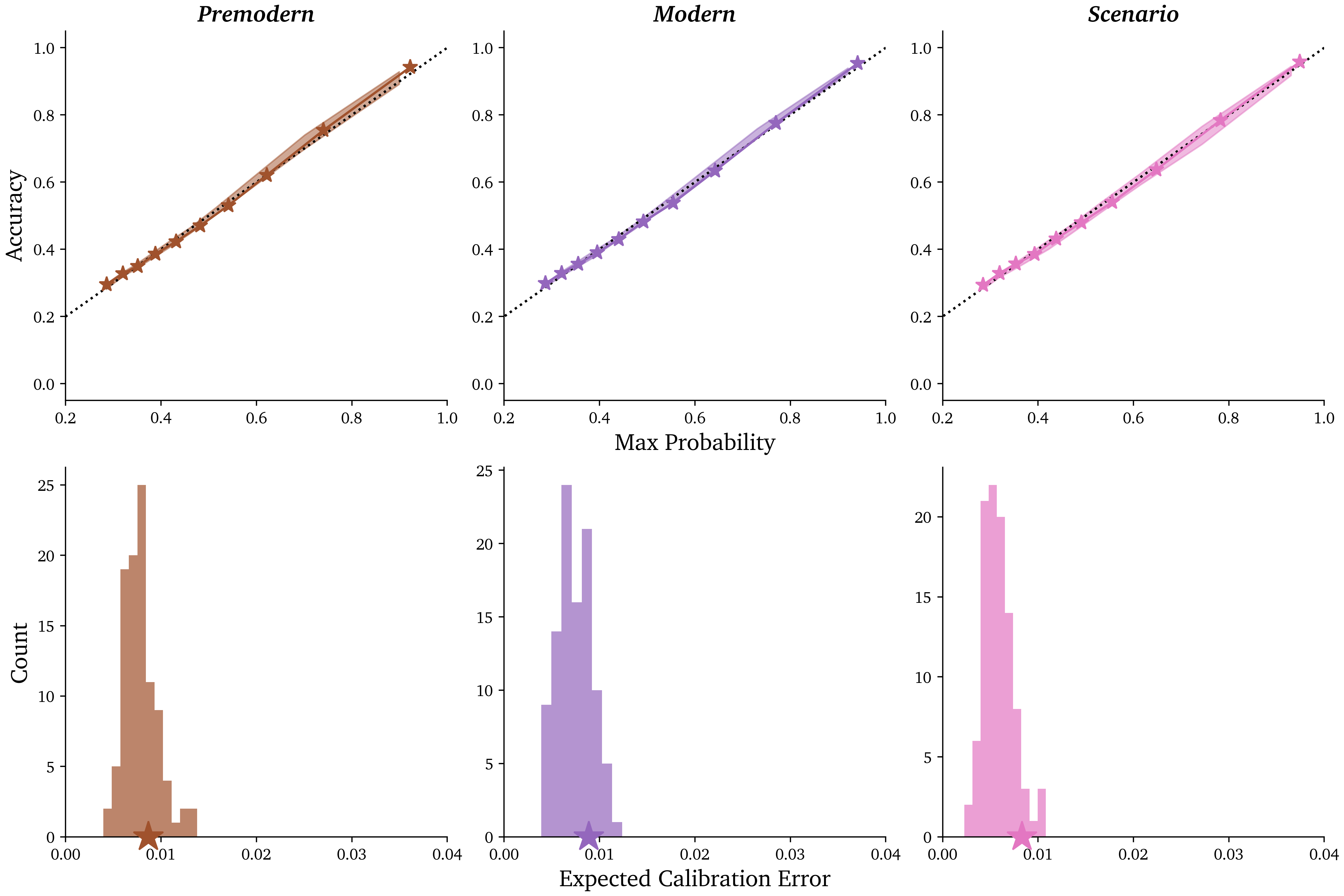}
     \caption{As in \fref{reliability}, but for ensembles with temperature scaling.}
    \label{reliability_scaled}
\end{figure}

\begin{figure}
    \centering
     \includegraphics[width=0.85\textwidth]{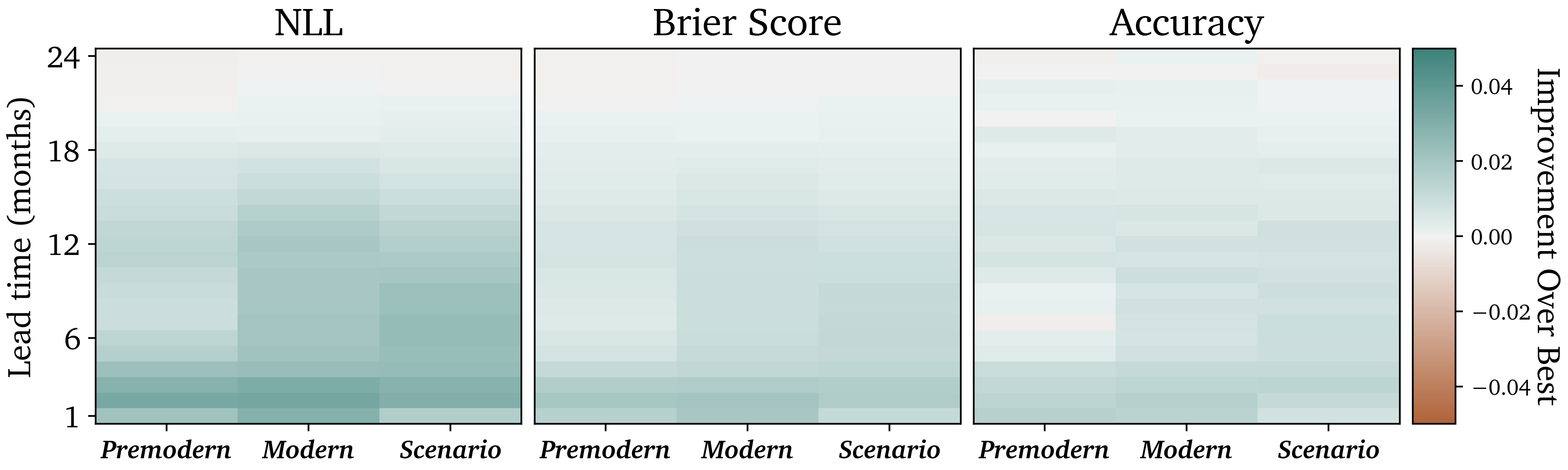}
     \caption{As in \fref{ens_vs_best}, but for ensembles with temperature scaling.}
    \label{ens_vs_best_scaled}
\end{figure}

Fortunately, deep ensemble underestimation bias can be effectively corrected post hoc through temperature scaling, which is the standard approach for correcting overestimation bias in overfit single models \parencite{guo}. Temperature scaling traditionally operates on logits, but can be implemented for ensemble probabilities, as in \textcite{rahaman}, by applying a simple scaling function $S$ to each probability vector $\bm{p}$ predicted by an ensemble. 

\begin{equation*}
    S(\bm{p}, T) = \frac{1}{Z}\left(({\bm{p}^{(1)}})^{\frac{1}{T}}, \dots, ({\bm{p}^{(K)}})^{\frac{1}{T}}\right)
\end{equation*}

where $Z$ is the normalizing constant and $T$ is the temperature. $T > 1$ increases the entropy of each predictive distribution and $T < 1$ decreases entropy. Since the underestimation bias is strongly influenced by lead time, we find a vector $\bm{T}$, containing a unique temperature for each lead time, which minimizes NLL on the validation set over the premodern period for each ensemble. We plot $\bm{T}$ for each ensemble in \fref{temps}. In \fref{reliability_scaled}, we show the resultant reliability diagrams after temperature scaling. For \fref{reliability_scaled}, we also temperature scale the logits of each component \parencite{guo}, as the softmax temperature implicitly learned in training can worsen component calibration biases \parencite{ashukha}.  Temperature scaling effectively mitigates the top-label underestimation bias, bringing the ensemble ECEs to within the distributions of component scores, as shown in \fref{reliability_scaled}. Importantly, after temperature scaling of both ensembles and components, the aforementioned short-range performance deficit in NLL and Brier score is corrected, as shown in \fref{ens_vs_best_scaled}. 

\clearpage
\section{XAI}
\label{sec:xai}

\begin{figure}[h!]
    \centering
     \includegraphics[width=1\textwidth,trim=4 4 4 4, clip]{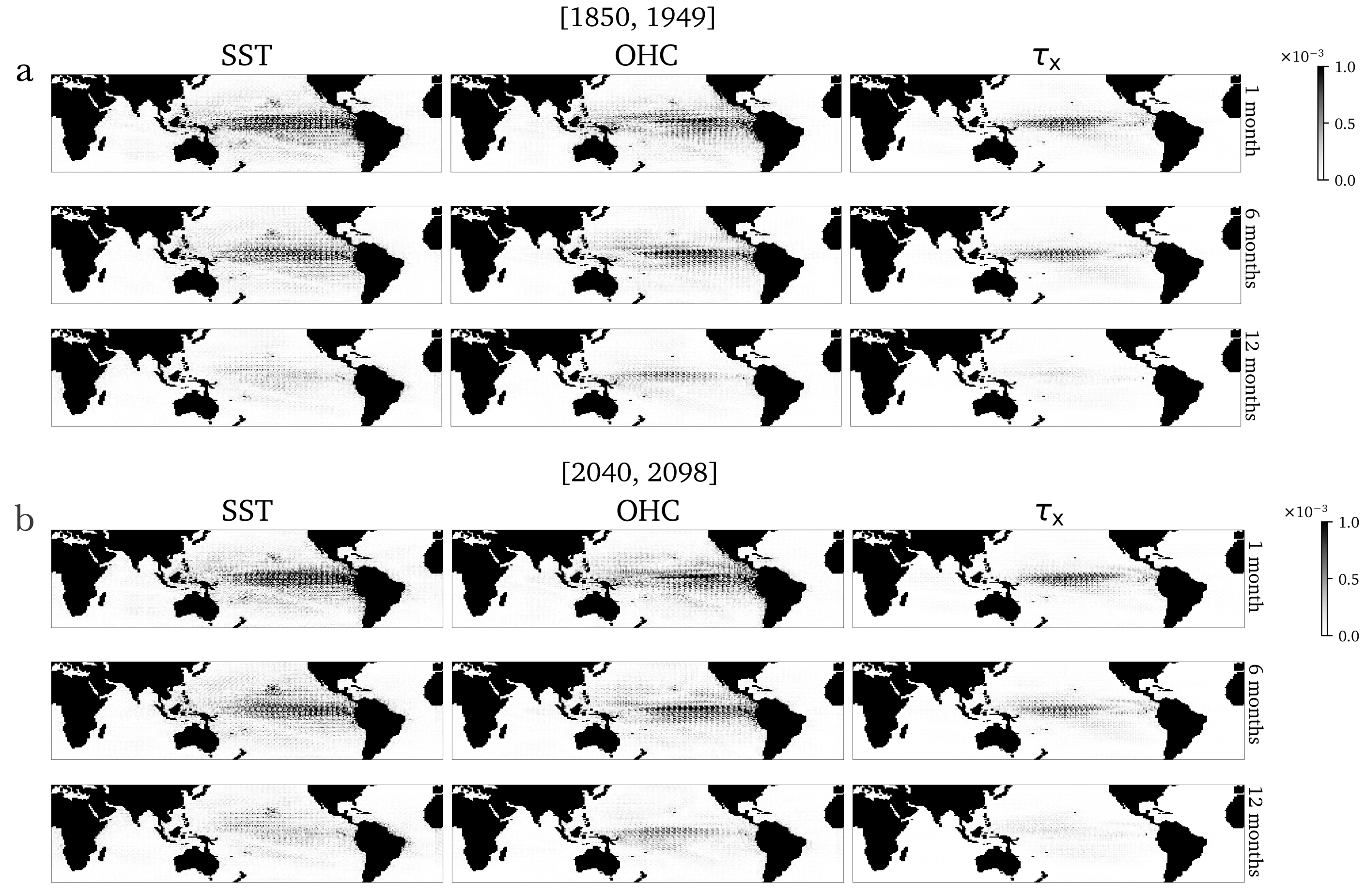}
     \caption{The absolute value of IG relevance averaged over five components of $\pmodern$ for the (a) premodern and (b) shifted periods. }
    \label{att}
\end{figure}

Integrated Gradients (IG) is an explainable artificial intelligence (XAI) technique, which estimates the importance of input features in determining neural network output \parencite{sundararajan}. IG quantifies the relevance of a local anomaly by integrating the gradient of the predicted class logit with respect to the local anomaly along a straightline path from a baseline value to the anomaly. We prescribe a zero baseline, i.e., climatology, and compute IG heatmaps for all testing set predictions of five components of $\pmodern$ for leads 1, 6, and 12 months. The composite heatmaps in \fref{att} suggest that, unsurprisingly, for both the premodern and shifted periods, components primarily focus on anomalies in the tropical Pacific when determining their predictions. 

\clearpage

\section{Additional Results}

\begin{figure}[h!]
    \centering
     \includegraphics[width=0.8\textwidth]{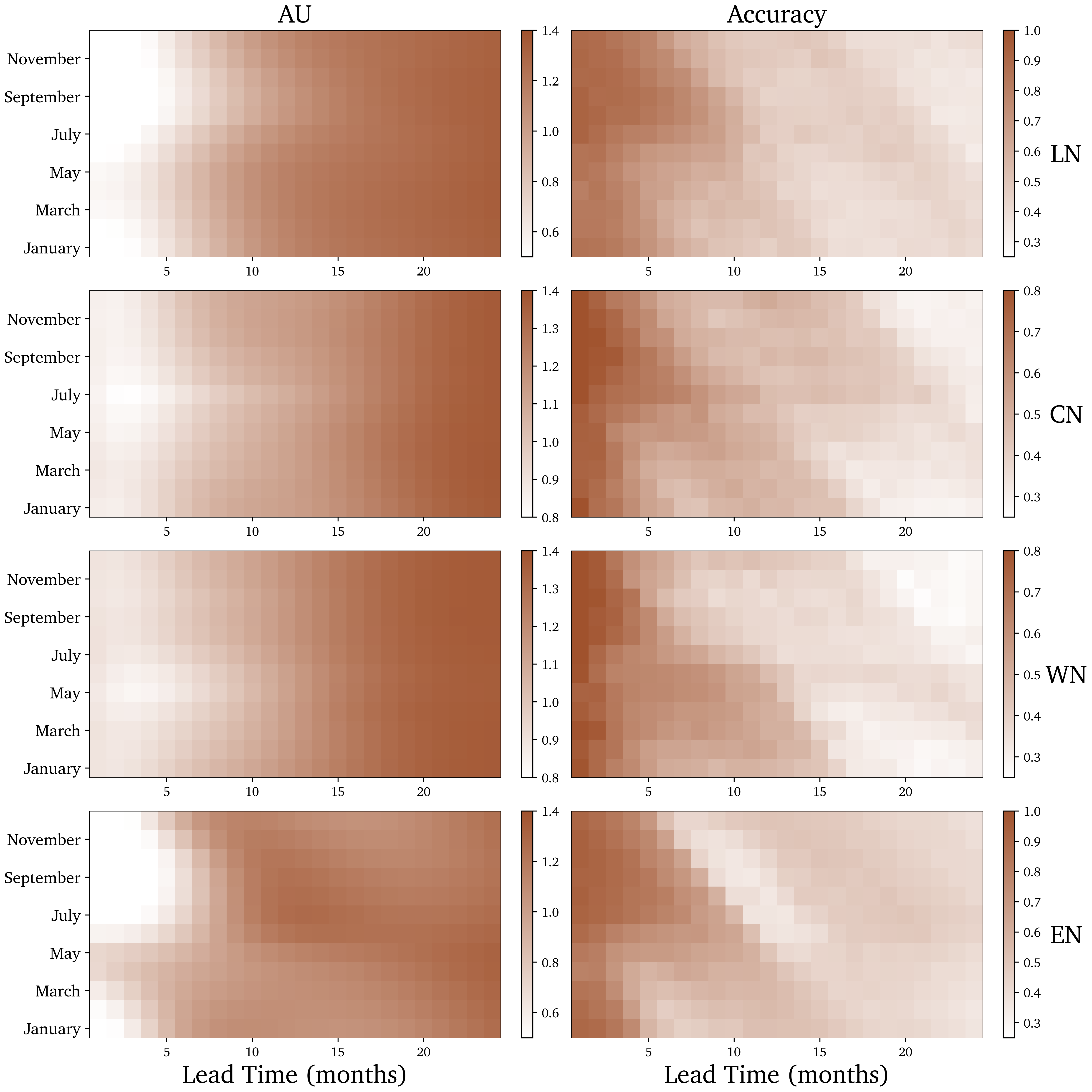}
     \caption{As in \fref{seasonality} separated on input phase for $\pmodern$. The SPB is muted but remains for CN and WN inputs, as shown in the accuracy plots. However, AU becomes less sufficient at signaling the spring predictability barrier for neutral conditions, especially for WN inputs, suggesting model deficiencies are contributing to the observed SPB.}
    \label{seasonality_phase}
\end{figure}

\begin{figure}[h!]
    \centering
     \includegraphics[width=0.7\textwidth]{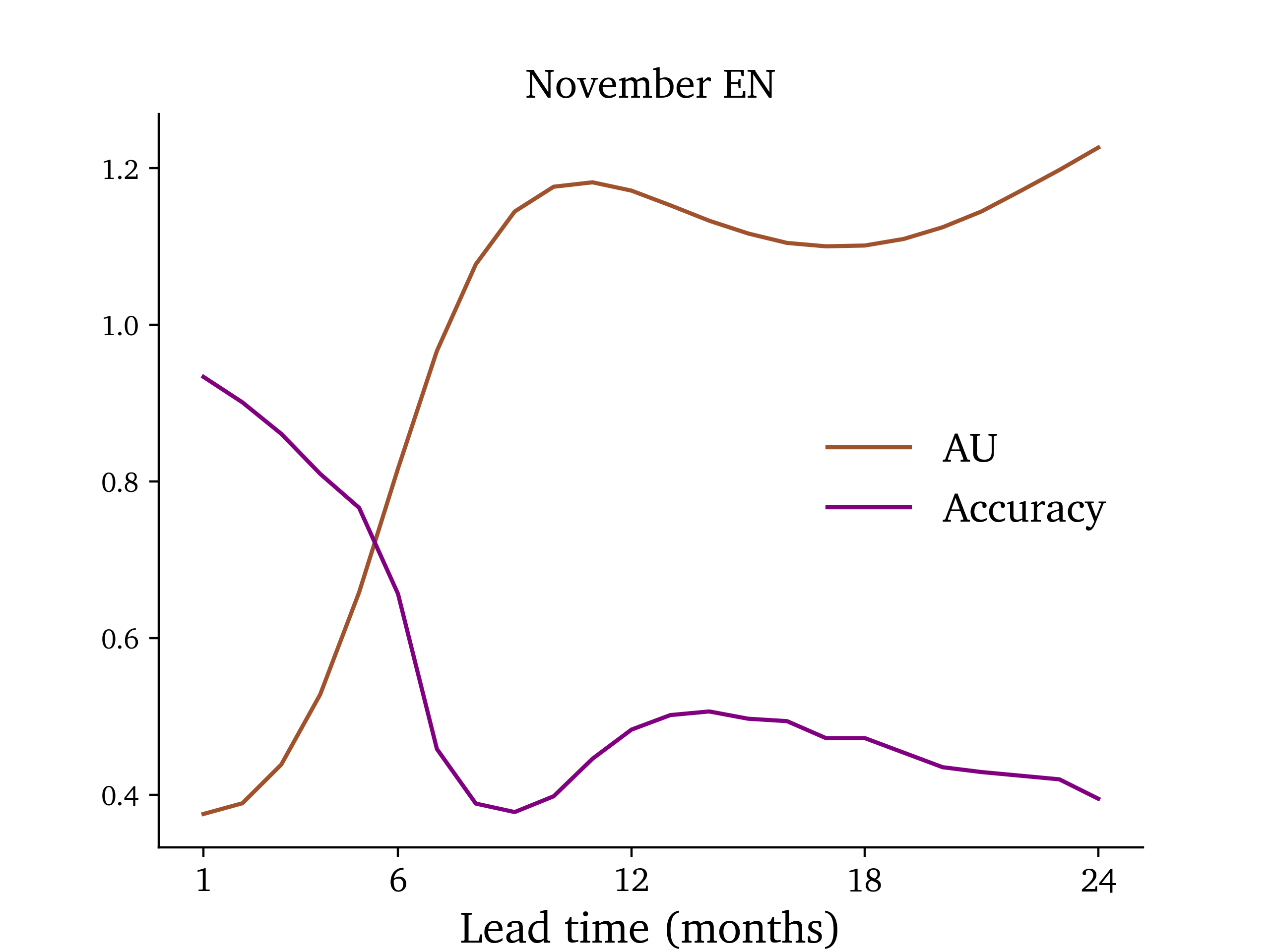}
     \caption{Mean AU and accuracy for $\pmodern$ ensemble for EN initializations during November.}
    \label{nov_en_au_acc}
\end{figure}

\begin{figure}[h!]
    \centering
     \includegraphics[width=0.6\textwidth]{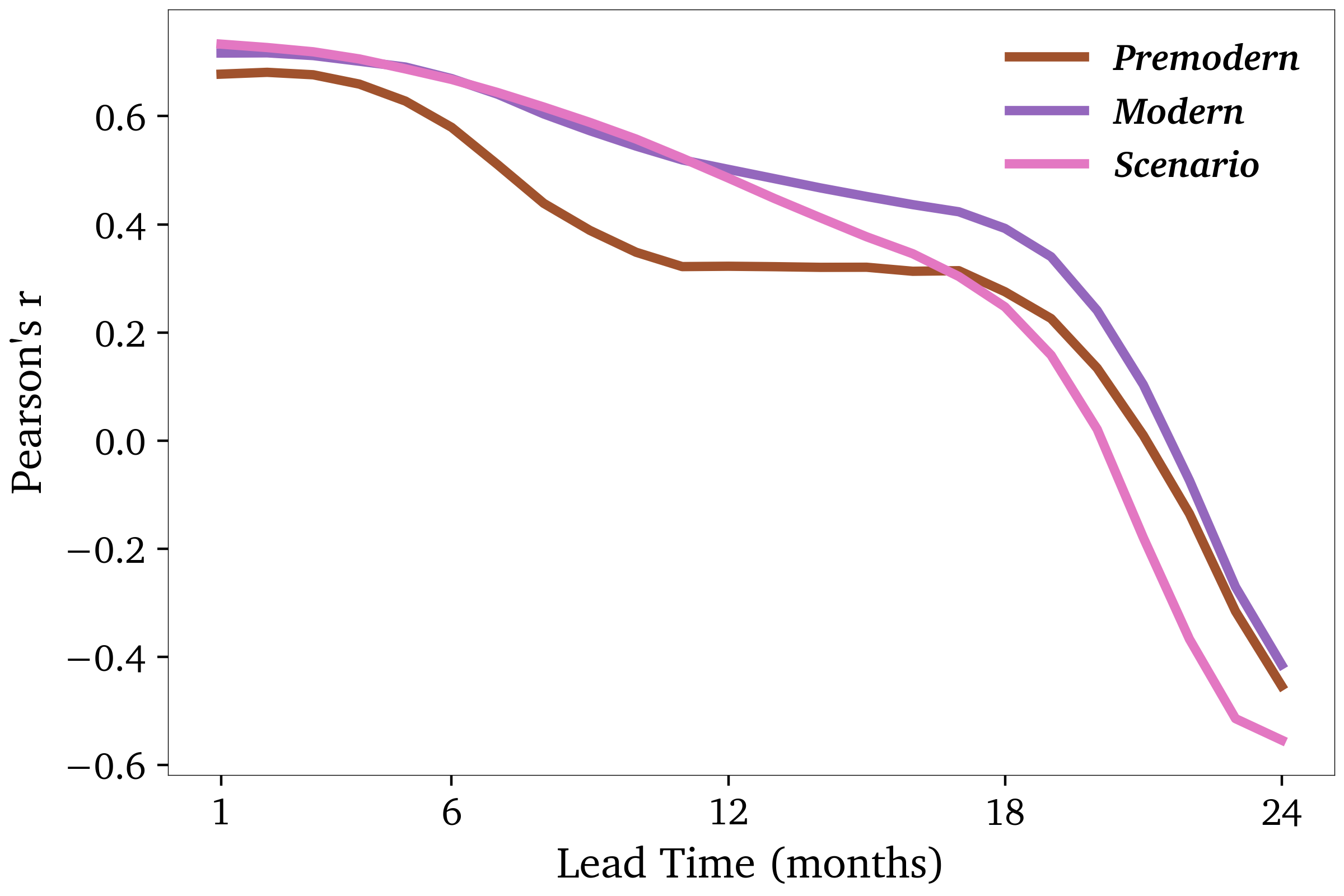}
    \caption{Linear correlation between AU and EU as a function of lead time for the testing set. The correlation is positive at short ranges, and is negative for the least predictable leads. Part of this behavior follows naturally from the definitions of AU and EU. For long leads, where mean AU is large, increasing AU concentrates ensemble members near the center of the probability simplex, decreasing EU. Whereas, for short leads, where mean AU is low, EU increases when certain components drift towards the center of the simplex away from the component-mean prediction, which also increases AU.}
    \label{au_eu_corr}
\end{figure}

\ifincludebib
    \newpage
    \printbibliography
\fi

\end{document}

        \newpage
        \printbibliography
    \end{refsection}
\fi

\end{document}